\documentclass[conference]{IEEEtran}
\usepackage{marvosym}

\usepackage{amsmath,amssymb,amsfonts}
\usepackage{algorithm}
\usepackage{algpseudocode}
\usepackage{graphicx}
\usepackage{textcomp}
\usepackage{subcaption}
\usepackage{xcolor}
\usepackage{fancyhdr}
\usepackage{cite}

\usepackage{enumitem}
\usepackage{lipsum}
\usepackage{booktabs}
\usepackage{multirow}
\usepackage{diagbox}
\PassOptionsToPackage{hyphens}{url}
\usepackage{hyperref}
\usepackage{tikz}
\usepackage{filecontents}
\usepackage{tabularx}
\usepackage{makecell}
\usepackage[absolute,overlay]{textpos}

\renewcommand{\baselinestretch}{0.97}

\newcommand{\distance}{1pt}
\newcommand{\incode}[1]{\lstinline{#1}}

\newcommand{\blue}[1]{\textcolor{black}{#1}}
\newcommand{\green}[1]{\textcolor{black}{#1}}

\newcommand{\eg}{e.g., }

\usepackage{xspace}
\newcommand{\sys}{\textsc{AdGPE}\xspace}
\newcommand{\ig}{\textsc{Pig}\xspace}

\newcommand{\eat}[1]{}

\usepackage{listings}
\definecolor{mygreen}{rgb}{0,0.6,0}
\definecolor{mygray}{rgb}{0.5,0.5,0.5}

\ifCLASSINFOpdf
\else
\fi

	\makeatletter
	
	\newcommand{\Rmnum}[1]{\expandafter\@slowromancap\romannumeral #1@}
	\makeatother
\begin{document}
%
\title{ Careful About What App Promotion Ads Recommend! Detecting and Explaining Malware Promotion via App Promotion Graph}

\author{\IEEEauthorblockN{Shang Ma\IEEEauthorrefmark{2}\textsuperscript{*},
Chaoran Chen\IEEEauthorrefmark{2}\textsuperscript{*},
Shao Yang\IEEEauthorrefmark{3},
Shifu Hou\IEEEauthorrefmark{2},\\
Toby Jia-Jun Li\IEEEauthorrefmark{2},
Xusheng Xiao\IEEEauthorrefmark{4}\Letter,
Tao Xie\IEEEauthorrefmark{5},
Yanfang Ye\IEEEauthorrefmark{2}\Letter}
\IEEEauthorblockA{\IEEEauthorrefmark{2}University of Notre Dame. Email: \{sma5, cchen25, shou, jli26, yye7\}@nd.edu}
\IEEEauthorblockA{\IEEEauthorrefmark{3}Case Western Reserve University. Email: sxy599@case.edu}
\IEEEauthorblockA{\IEEEauthorrefmark{4}Arizona State University. Email: xusheng.xiao@asu.edu}
\IEEEauthorblockA{\IEEEauthorrefmark{5}Peking University. Email: taoxie@pku.edu.cn}
}



\IEEEoverridecommandlockouts
\makeatletter\def\@IEEEpubidpullup{6.5\baselineskip}\makeatother
	\IEEEpubid{\parbox{\columnwidth}{\rule{\columnwidth/2}{0.5pt}\\
    $^{\textrm{\Letter}}$ Corresponding authors. * Equal contribution.\\ \\
    Network and Distributed System Security (NDSS) Symposium 2025\\
    24-28 February 2025, San Diego, CA, USA\\
    ISBN 979-8-9894372-8-3\\
     https://dx.doi.org/10.14722/ndss.2025.230051\\
    www.ndss-symposium.org
}
\hspace{\columnsep}\makebox[\columnwidth]{}}

\maketitle

\begin{abstract}
In Android apps, their developers frequently place app promotion ads, namely advertisements to promote other apps. 
Unfortunately, the inadequate vetting of ad content allows malicious developers to exploit app promotion ads as a new distribution channel for malware.
%

To help detect malware distributed via app promotion ads, in this paper, we propose a novel approach, named \sys, that synergistically integrates app user interface (UI) exploration with graph learning to automatically collect app promotion ads, detect malware promoted by these ads, and explain the promotion mechanisms employed by the detected malware. 

Our evaluation on $18,627$ app promotion ads demonstrates the substantial risks in the app promotion ecosystem. The probability for encountering malware when downloading from app promotion ads is hundreds of times higher than from the Google Play. Popular ad networks such as Google AdMob,
Unity Ads, and Applovin are exploited by malicious developers to spread a variety of malware: aggressive adware, rogue security software, trojan, and fleeceware.
Our UI exploration technique can find $24\%$ more app promotion ads within the same time compared to the state-of-the-art techniques.
We also demonstrate our technique's usage in investigating underground economy by collecting app promotion ads in the wild.
Leveraging the found app promotion relations, our malware detection model achieves a $5.17\%$ gain in F1 score, improving the F1 score of state-of-art techniques from $90.14\%$ to $95.31\%$.
Our malware detection model also detects $28$ apps that were initially labeled as benign apps by VirusTotal but labeled by it as malware/potentially unwanted apps (PUAs) six months later. Our path inference model unveils two malware promotion mechanisms: custom-made ad-based promotion via hardcoded ads and ad library-based promotion via interactions with ad servers (\eg AdMob and Applovin).
These findings uncover the critical security risks of app promotion ads and demonstrate the effectiveness of \sys in combining dynamic program analysis with graph learning to study the app promotion ad-based malware distribution.
\end{abstract}


%

\section{Introduction}

Advertisements, in short as ads, are widely used in mobile apps.  
Research~\cite{viennot2014measurement} indicates that over $57\%$ of apps in Google Play contain ad libraries. 
Additionally, two-thirds of popular apps contain these ad libraries~\cite{liu2015efficient}.
Ads are effective and widely adopted practices to increase user bases and app installs, and are also a major revenue source for developers who place ads in their apps to promote other apps (denoted as app promotion ads in this paper)~\cite{applovinsuccessstories,admobadvantages}.
These ads play a significant role in helping users discover new apps.
For example, a survey~\cite{GoogleResearch} from Google Research and Ipsos shows that one-third of users discover new apps via ads in other apps.
Unfortunately, the inadequate vetting of the ad content allows malicious developers to exploit app promotion ads as a new distribution channel for malware. 





\begin{figure}
    \centering
    \includegraphics[width=0.95\columnwidth]{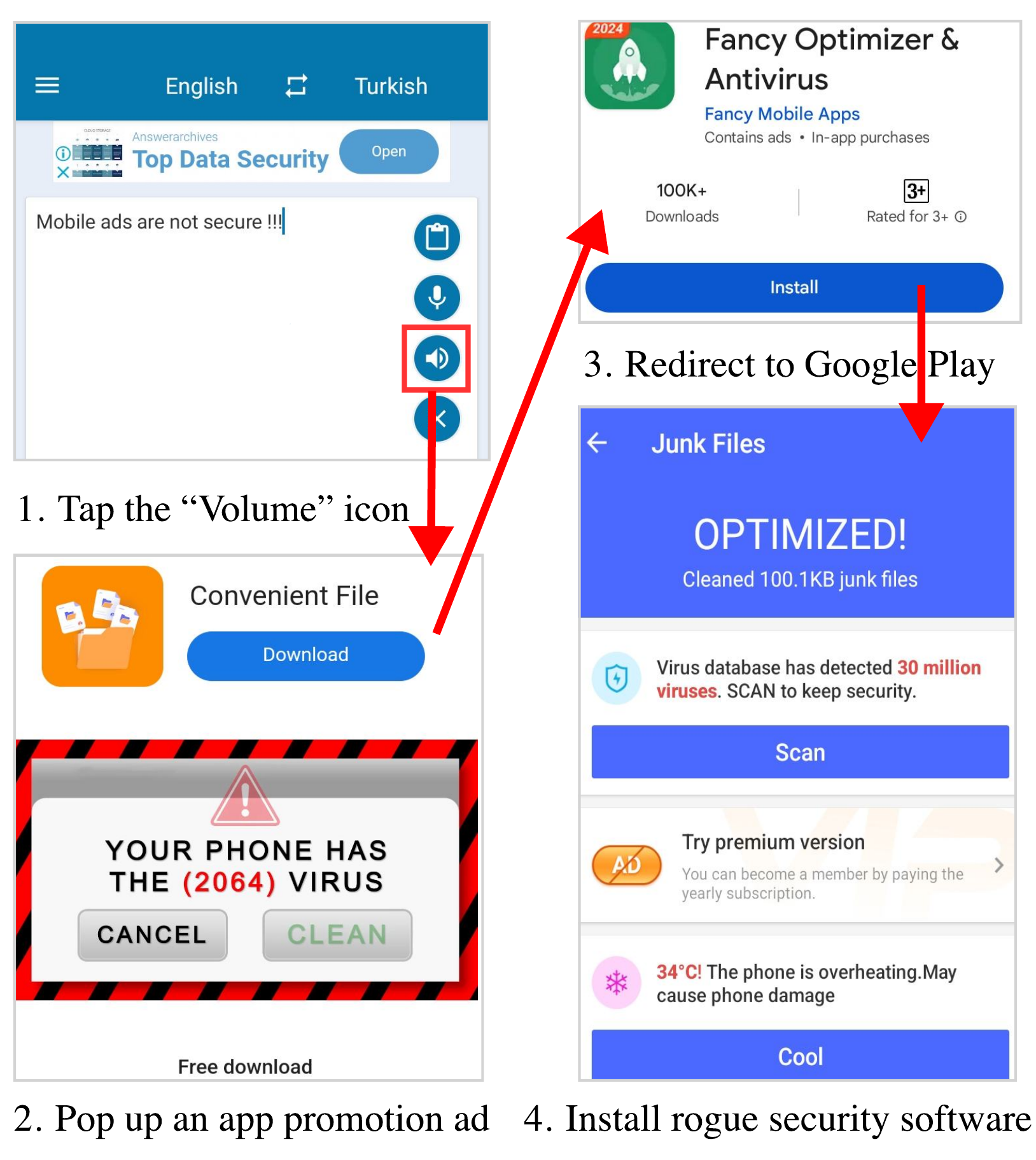}
    \vspace*{-1ex}
    \caption{Malware distribution via an app promotion ad}
    \label{fig: app promotion ad example}
\end{figure}

%

App users are prone to malware promotion ads for two main reasons.
First,  these ads are usually ``hidden'' in harmless apps, engaging users with ad contents. 
For example, the app shown in Fig.~\ref{fig: app promotion ad example}  offers free translation services between English and Turkish. 
Nevertheless, any interaction with this app, such as tapping the ``Voice'' icon, will trigger an app promotion ad.
This ad employs persuasive language and visual patterns to encourage users to download rogue security software that convinces users to pay for security services and steals users' sensitive information.
 Second,  the malware promotion ads are usually indistinguishable from other ads, leading users to unknowingly download malware from trusted sources.
In Fig.~\ref{fig: app promotion ad example}, tapping anywhere on the ad redirects users to the official Google Play store to install the advertised app.
Moreover, this ad is provided by Google AdMob, which is the most widely used ad library, integrated into over $90\%$ of apps that conduct app promotion campaigns. 
Similar malware promotion ads are also observed from other popular ad libraries such as Meta Ads and AppLovin.



To help detect malware distributed via app promotion ads, in this paper, we propose a novel approach, \sys, that \textbf{synergistically integrates app UI exploration with graph learning to automatically collect app promotion ads, detect malware promoted by these ads, and explain the promotion mechanisms employed by the detected malware}\footnote{\sys is publicly available at \url{https://github.com/AppPromotionAdsResearch/AdGPE}.}.
\blue{In particular, while static analysis can effectively identify ad libraries used to show ads in apps~\cite{ma2016libradar,li2017libd,liu2020maddroid}, our preliminary study finds that applying static analysis to detect app promotion ads is insufficient because (a) the ad content served by ad libraries is determined at runtime and dynamically requested from ad servers, and (b) custom-made ads (ad content provided by the app developer instead of ad libraries) that account for non-trivial portions of ads ($>20\%$) have diverse implementation mechanisms
(Section~\ref{subsec:characteristics_ads}).}
Thus, to address these challenges, \sys performs automatic user interface (UI) exploration on a large number of apps from Google Play to collect app promotion ads, and leverages the collected ads to build \textit{\textbf{an app promotion graph}}.
\blue{In an app promotion graph, the nodes represent apps, and the edges represent the relationships from the apps showing the ads to the apps promoted by the ads.}
This graph encapsulates the app promotion relations among apps and the app attributes derived from app markets, VirusTotal, and source code. 
Utilizing the graph, \sys then trains a node classification model for malware detection and constructs a path inference model to uncover malware promotion mechanisms.

\sys starts by collecting app promotion ads to construct an app promotion graph. 
We have pinpointed three major challenges of this task.
First, app promotion ads are often placed on multiple UI pages within an app. 
Existing UI exploration techniques aim at achieving high code coverage within a single app by employing complex and heavy strategies, which become inefficient when scaling to a large number of apps~\cite{gao2018android,mirzaei2012testing,guitest1,guitest2,guitest3}.
Second, app promotion ads appear in various formats, each with distinct UI layout characteristics. 
Current works on identifying mobile ads are limited to a small set of UI widgets thus failing to capture all the app promotion ads~\cite{liu2020maddroid,nath2015madscope,rastogi2016these}.
Third, ads’ content is generated at runtime
and continuously changes over time.
Existing UI exploration techniques are optimized on code coverage so they tend to avoid repeated actions and may overlook the changed ad content~\cite{hu2011automating,nath2013smartads,liu2014decaf,hao2014puma}. 
To address these three challenges, \sys includes a novel ad-oriented UI exploration technique. 
First, to efficiently achieve high ad coverage, this technique employs random exploration combined with a depth-first exhaustive search strategy.
Second, to accurately identify various ad types, this technique leverages a common design feature: a call-to-action widget redirecting users to the app marketplace.
Third, to capture the frequently updated ad content, this ad-oriented UI exploration technique periodically restarts apps and refreshes their ads to capture all available content within a period of time.

The second part of \sys is to leverage the constructed app promotion graph for ad-promoted malware detection.  
Existing Android malware detection models have two shortcomings when detecting ad-promoted malware~\cite{bordes2013translating,hou2017hindroid,zhang2020enhancing}.
First, they are ineffective as they mainly train learning models using only app attributes (\eg permissions, app metadata, and API calls) and \textit{overlook the app promotion relations}.
Second, these models generally \textit{lack explainability}. 
To address these shortcomings, \sys introduces a novel graph learning model based on pre-trained Graph Neural Network (GNN)~\cite{kipf2016semi, hamilton2017inductive, xu2018powerful, hu2019strategies}.
To effectively detect ad-promoted malware, this model constructs an embedding of the app promotion graph to represent app promotion relations.
This embedding is then combined with the app attributes to train a Random Forest classifier~\cite{ho1995random} for malware detection.
To enhance the explainability of our model, \sys transforms the app promotion graph into a promotion inference graph (\ig) and builds a path inference model to infer malware promotion mechanisms based on app attributes and app promotion relations. 
The path inference model also predicts unobserved links in the app promotion graph, complementing UI exploration to build a more comprehensive app promotion graph for better detection and inference.

Our study on $18,627$ app promotion ads uncovers significant risks in these ads: a $2.64\%$ chance for users to encounter malware when downloading apps via app promotion ads, which rises to $7.51\%$ when the ads are in potentially unwanted apps (PUAs).
This probability is \textit{\textbf{hundreds of times higher than the likelihood of downloading a malicious app from the Google Play store}}.
\blue{Furthermore, our case study reveals that \textbf{\textit{popular ad libraries such as Google AdMob, Unity Ads, and Applovin lack stringent vetting processes for malware and are exploited by malicious developers to spread a variety of malware: aggressive adware, rogue security software, trojan, and fleeceware}}. App developers can also build custom-made ads to distribute malware. Although app markets have strict policies, they alone cannot mitigate all risks without cooperation from ad libraries and developers. Thus, solely analyzing ad libraries is not sufficient.}

By incorporating app promotion relations,  our malware detection model 
 obtains a \textbf{\textit{ $5.17\%$ performance gain}} (from $90.14\%$ to $95.31\%$) compared to using solely traditional features\footnote{In this paper, the term ``malware'', when used in the context of malware detection or malware promotion, encompasses both the malware and PUAs.}.
\blue{This indicates that the app promotion graph built through UI exploration is an important feature that enhances the effectiveness of graph learning in malware detection.}
\blue{Specifically, it outperforms five commercial security engines by at least $27.4\%$ and two state-of-the-art (SOTA) malware detection approaches~\cite{onwuzurike2019mamadroid,xu2019droidevolver} by at least $24.5\%$ in F1 score.
Additionally, \sys achieves a better F1 score than the SOTA approach that employs random walk~\cite{shen2021andruspex} and is fed with the same features of \sys ($95.31\%$ v.s. $92.48\%$), indicating the superiority of \sys's GNN model.
We evaluate the robustness of \sys by mimicking real-world attacks through mutating nodes and links in the app promotion graph. 
The results demonstrate that \sys maintains a high F1 score with node mutations and outperforms previous work with link mutations.}
Our malware detection model also successfully detects $28$ apps that were initially labeled as benign apps by VirusTotal but labeled as malware/PUAs six months later. 
\blue{Furthermore, our ad-oriented UI exploration technique outperforms four established approaches~\cite{li2017droidbot,monkey,liu2020maddroid,cai2023darpa}, by finding $24\%$ more ads within $10$ hours.}
Our small-scale case study on collecting app promotion ads in the wild also shows that this UI exploration technique can effectively detect ads directing users outside Google Play and discover malware from non-official app markets, assisting in studying underground economy.
Our path inference model reveals two primary malware promotion channels:  ad library-based promotion via interactions with ad servers (\eg AdMob, Applovin) and custom-made ad-based promotion via hardcoded ads (\eg apps from the same malicious developer) in the app's source code.
This path inference model also aids ad-oriented UI exploration by predicting missing app promotion links, some of which require at most $54$ clicks to find. 
These findings  demonstrate \textbf{\textit{  the effectiveness of combining dynamic program analysis with graph learning in studying app promotion ad-based malware distribution.}}

\section{Background and Threat Model}
\noindent\textbf{Mobile Advertisement Ecosystem.}
\label{sec: mobileAdsEcosys}
\green{The mobile advertising ecosystem includes several key roles: \textit{Ad Networks}, which conduct real-time auctions to determine which ads to display based on factors such as bid price, ad quality, and user context~\cite{googleadsAuction,facebookadsAuction}; \textit{Ad Providers}, who aim to promote their apps via ads; and \textit{Developers}, who seek to monetize their apps by hosting ads.
Ad providers must publish their apps on an app market and register these apps with an ad network to finance advertisement campaigns \cite{googleAds}.  Ad networks, which are incorporated into mobile apps via ad libraries ~\cite{michael2016adlib,erik2017adlib},  support various formats like banners and rewarded ads to enhance user visibility~\cite{admob_adformats,facebook_adformats,applovin_adformats}. 
To host ads, developers need to register their apps with an ad network and integrate the network's ad library into their apps~\cite{admobGetStarted}.
Additionally, to reduce advertising expenses, developers can create custom-made ads within their own apps to promote other products or apps they offer.}

\noindent\textbf{Automatic UI Exploration.}
Automatic UI exploration for Android devices is an automated approach that systematically navigates and interacts with an app's UI to identify potential issues, validate functionality, and ensure a seamless user experience~\cite{choudhary2015automated, kong2018automated}. 
Recognizing the importance, Google has released the UI exerciser, the Monkey tool~\cite{monkey}, which is a command-line tool to randomly generate user events including clicks, touches, and gestures.
There also exist research tools that model the explored UI pages and transitions to automatically explore various behaviors of apps~\cite{guitest1,guitest2,guitest3,paladin}.
Apart from these automated UI testing tools, there are also UI testing techniques that enable developers to create customized code to automate the UI tests, such as Appium~\cite{Appium} and Google's UIAutomator~\cite{uiautomator}.

\noindent\textbf{Threat Model.} We conduct the study in our controlled lab environments with trusted devices and secure network settings, and our threat model is consistent with the previous work~\cite{son2016mobile, chen2014peeking}. 
Under this lab setting, we assume that app promotion ads in our experiments are not subject to external attacks.
Specifically, we assume that the ads are generated by the apps themselves instead of from other apps that preempt the ad promotion channel to distribute ads in the other apps. 
We also assume that the network channels and the ad servers of the apps are not compromised and the ad content is not altered. 
Although such scenarios may occur, they fall outside the scope of this research. 
To ensure that the annotation of the app class (i.e., benign app, PUA, and malware) is accurate and our trained model is not biased or polluted, 
we not only rely on the VirusTotal security report of each app, 
but also sample and inspect the source files of the reported malware to ensure the existence of malicious behavior in the code.
Malware that deliberately evades detection and inspection can be detected by applying more advanced techniques~\cite{genome,appverify,appcontext,deepintent,deeprefiner}, which is out of the scope of this paper.

\section{Motivation of \sys}

\begin{figure}
    \centering
    \includegraphics[width=\columnwidth]{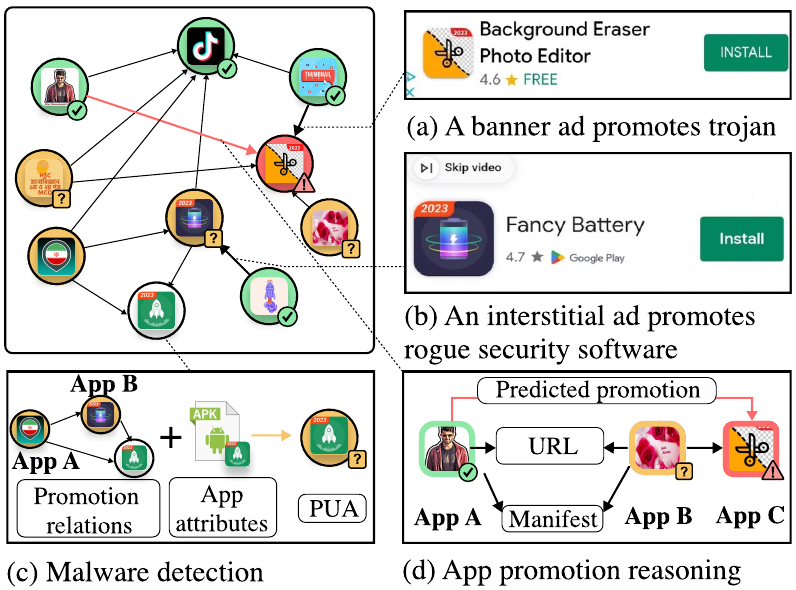}
    \caption{A motivating example}
    \label{fig: motivation_example}
\end{figure}

In this section, we introduce the motivation behind our approach, starting with an example that highlights the potential risks associated with app promotion ads. 
We also use an example to illustrate the benefits of integrating graph learning with UI exploration, demonstrating the enhanced effectiveness of this synergistic approach.

\subsection{Risks in App Promotion Ads}

Fig.~\ref{fig: motivation_example} shows a real-world app promotion graph where apps are connected by app promotions ads (arrows pointing to promoted apps). 
This graph consists of malware (the icon with a red triangle), PUAs (icons with a yellow rectangle), and benign apps (icons with a green circle).
The malware ``Background Eraser Photo Editor'' and the PUA ``Fancy Battery'' can both be accessed by users following the app promotion ads seen in Fig.~\ref{fig: motivation_example}(a) and (b), both of which are located in benign apps. 
The malware ``Background Eraser Photo Editor'' is a trojan that disguises itself as a photo editor while executing malicious activities such as stealing sensitive information and e-banking frauds. 
The PUA ``Fancy Battery'' is a rogue security engine that disguises itself as a cleaner app but executes malicious services automatically to constantly display ads without requiring user interactions.


Beyond leveraging ad libraries, malware is also 
promoted through custom-made app promotion ads.
Fig.~\ref{fig:magnet_customized} in the Appendix shows an example app that pops up a dialog in a benign app and directs users to download a trojan.
\textit{Clearly, these examples show that malware/PUAs can exploit app promotion ads in benign apps, which are trusted by most users, to reach more users.}

\subsection{Graph Learning with App Promotion}

Fig.~\ref{fig: motivation_example}(c) presents an example of using app promotion relations to detect malware.
 Specifically, the adware ``Learn Persian", which promotes the target app, ``Fancy Boost", also promotes another PUA named ``Fancy Optimizer \& Antivirus", both developed by the same developer. 
 Such app promotion relations suggest a high likelihood that the target app belongs to the same group of malicious developers (referring to Section~\ref{sec: mal_promote} for more details).
The target app ``Fancy Boost" indeed exhibits unwanted behaviors similar to the rogue security software ``Fancy Optimizer \& Antivirus" (as shown in Fig.~\ref{fig: app promotion ad example}).
Nevertheless, this app is a newly developed app with limited information and security vendors such as VirusTotal classify it as benign.  
This example shows how app promotion relations facilitate graph learning-based malware detection by providing distinctive features.

Fig.~\ref{fig: motivation_example}(d) shows how to use an app promotion graph to predict and reason app promotions. 
Specifically, the online message sticker app ``Vijay Deverakonda Sticker" (referred to as App A) accesses the same ad library URL and shares manifest activity names with the photo collection app ``Morning and Night Wishes" (referred to as App B), which is known to promote the malware ``Background Eraser Photo Editor" (referred to as App C).
This information implies that App A and App B share exploit code and use the same ad library, suggesting a strong likelihood of App A promoting App C. 
However, real-time UI exploration may not capture every instance of app promotion ads, potentially missing this malware promotion.
This example demonstrates that graph learning offers a solution by predicting app promotions, effectively complementing dynamic analysis to address its inherent limitations, such as the incompleteness of app UI exploration.

\subsection{Characterizing Challenges for App Promotion Ads}
\label{subsec:characteristics_ads}

\begin{figure}[t!p]
	\begin{lstlisting}
import com.google.android.gms.ads.AdView;
import com.google.android.gms.ads.InterstitialAd;
// Inherent ads
AdView adView = findViewById(R.id.adView); 
adView.loadAd(ConsentSDK.getAdRequest(this));
// Pop-up ads
InterstitialAd popupad = new InterstitialAd(this);
popupad.loadAd(ConsentSDK.getAdRequest(this));
adButton.setOnClickListener(view -> popupad.show()); 
// Custom-Made ads
customAdButton.setOnClickListener(view -> startActivity(new Intent(Intent.ACTION_VIEW, Uri.parse(uri))));\end{lstlisting}
 \caption{\blue{Key code snippets of three ad types}}
	\label{fig:adview}
\end{figure}

  \begin{figure}[t]
    \begin{subfigure}[]{0.32\linewidth}
      \centering
      \includegraphics[width=\linewidth]{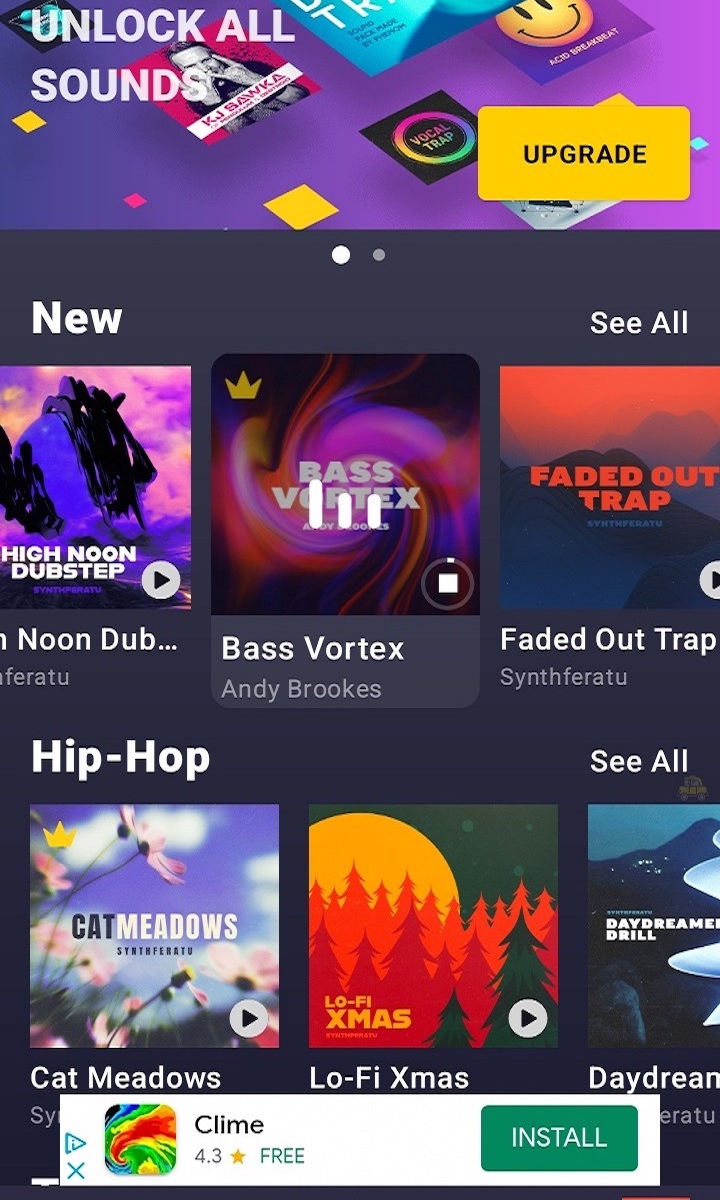}
      \caption{Inherent Ad}
      \label{fig:banner}
    \end{subfigure}
        \begin{subfigure}[]{0.32\linewidth}
      \centering
      \includegraphics[width=\linewidth]{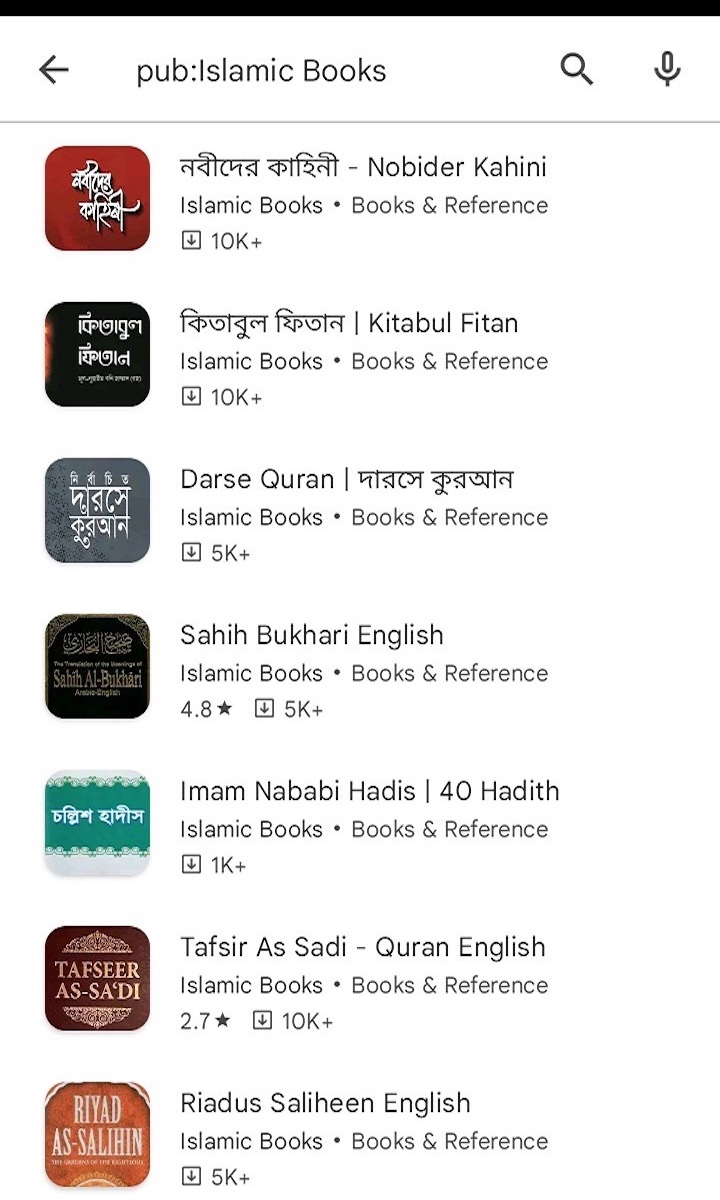}
      \caption{Custom-made Ad}
      \label{fig:more_apps}
    \end{subfigure}
    \begin{subfigure}[]{0.32\linewidth}
      \centering
      \includegraphics[width=\linewidth]{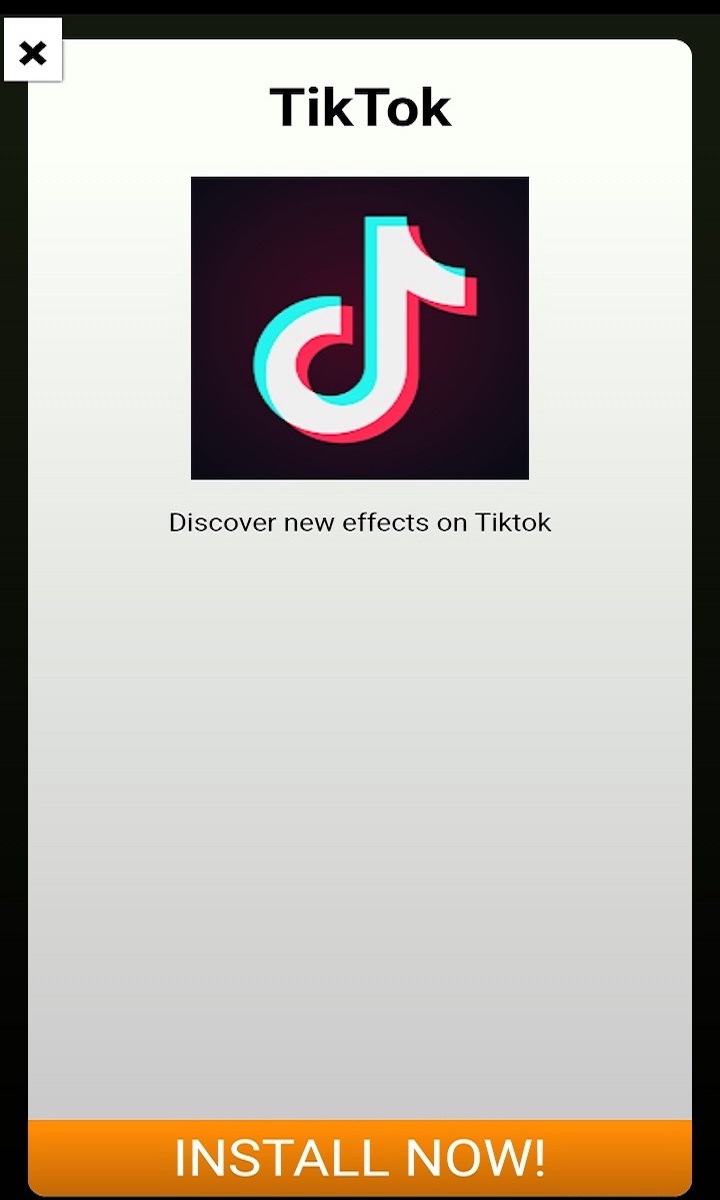}
      \caption{Pop-up Ad}
      \label{fig:interstitial}
    \end{subfigure}
     \caption{Examples of app promotion ads}
    \label{fig:ad_formats}
  \end{figure}

\blue{
App promotion ads are shown in various formats~\cite{admob_adformats,facebook_adformats,applovin_adformats}, including banner, interstitial, and rewarded ads.  
Moreover, app promotion ads can be implemented by \textit{ad libraries} or \textit{custom-made by developers}, posing challenges for collecting app promotion ads. 
Therefore, we conduct a preliminary study to investigate the characteristics of app promotions ads and how we can collect them. 
We have built two datasets as follows:}


\noindent\textbf{AndroZoo Dataset.} We first construct a dataset of advertiser apps based on AndroZoo~\cite{allix2016androzoo}, a comprehensive collection of APK files that is regularly updated and widely acknowledged in the research community.
Our selection criteria from AndroZoo are as follows: 1) the apps must have been released after 2020 to ensure that their advertising practices are current; 2) they must include ad libraries, indicating a likelihood of containing advertisements; 3) they should have a Play Store page to largely guarantee the app's quality;
4) their download in Play Store should be diversified, as apps with different popularity may have different advertising practices. 
We filter the dataset with these criteria and then randomly sample $200$ apps of the resulting dataset.
We run these apps on Android phones and manually explore these apps to record the app promotion ads and the app promoted by those ads.

\noindent\textbf{Rico Dataset.} Manually exploring app promotion ads in a large number of apps from AndroZoo requires extensive human labor.
To support a comprehensive study, we additionally use Rico dataset~\cite{deka2017rico}, which
provides screenshots, view hierarchies and human annotations.
 The screenshots and corresponding view hierarchies directly reveal the GUI information of app promotion ads, such as the attributes of the ad-related widget and the placement of the ad-related widget in the UI.
 Human annotations (including interaction traces) tell the location of ad-related UI in the whole app and how to reach it. 
Specifically, we refine this dataset to include only screenshots labeled as ``Advertisemen'', yielding $1326$ screenshots with view hierarchies from $405$ apps.
We then manually review these screenshots and view hierarchies.

\begin{table}[t!]
\caption{\blue{Dataset of the preliminary study}}
\centering
\label{tab:adtypes}
\begin{tabular}{@{}lrrrr@{}}
\toprule
\multirow{2}{*}{\textbf{Ad types}} & \multicolumn{1}{c}{\multirow{2}{*}{\textbf{\begin{tabular}[c]{@{}c@{}}Using ad\\ libraries\end{tabular}}}} & \multicolumn{2}{c}{\textbf{\# Samples}}                                   & \multicolumn{1}{c}{\multirow{2}{*}{\textbf{Example}}}       \\ \cmidrule(lr){3-4}
                          & \multicolumn{1}{c}{}                                                                                       & \multicolumn{1}{c}{\textbf{AndroZoo}} & \multicolumn{1}{c}{\textbf{Rico}} & \multicolumn{1}{c}{}                                        \\ \midrule
Inherent ads              & Yes                                                                                                        & 75                                    & 167                               & Fig.~\ref{fig:banner}                                     \\
Pop-up ads                & Yes                                                                                                        & 69                                    & 22                                & Fig.~\ref{fig:interstitial} \\
Custom-Made ads                & No                                                                                                         & 45                                    & 5                                 & Fig.~\ref{fig:more_apps}                                  \\ \bottomrule
\end{tabular}
\end{table}

\noindent\textbf{Categorization of Ad Types.}
\blue{
We browse the documentation of ad libraries~\cite{admob_adformats,facebook_adformats,applovin_adformats} and also manually decompile the advertiser apps to inspect the code logic triggering app promotion ads. 
Based on our observations,  we categorize the recorded ads into three ad types.
Fig.\ref{fig:adview} illustrates the key code snippets for each ad type, while Fig.\ref{fig:ad_formats} provides examples of these ad formats.
\begin{itemize}[noitemsep, topsep=1pt, partopsep=1pt, listparindent=\parindent, leftmargin=*]
    \item \textit{Inherent ads:} These ads are shown using visible views created via ad libraries (Line 1 and Line 4), and are usually shown when apps are started.
    \item \textit{Pop-up ads:} These ads are created using ad libraries (Line 2 and Line 7) and are usually triggered by user interactions, such as clicks (Line 9).
    \item \textit{Custom-Made ads:} These ads are developed by app developers and can be of various forms such as buttons, banners, or app walls. The ad content (such as the URI of the advertising site shown in Line 11) can be hardcoded or dynamically requested from developers' own servers.
\end{itemize}
}

\blue{
It can be seen that ad content from both inherent ads and pop-up ads are dynamically requested from the ad library via \incode{getAdRequest} (Lines 5 and 8), and thus cannot be analyzed by static analysis. 
It can also be challenging for static analysis to analyze custom-made ads due to the highly customized implementations from different developers. }

\noindent\textbf{Ad Characteristics for Malware Promotion.}
\blue{
Table~\ref{tab:adtypes} summarizes the characteristics of these ad types and their distribution in the two datasets.  
From the AndroZoo dataset, we can find that while ads from ad libraries (i.e., inherent and pop-up) are the majority, custom-made ads account for $23.81\%$ ($45/189$) of the ads, indicating that custom ads represent a non-trivial portion of the ads analyzed. 
Furthermore, we have scanned the promoted apps via VirusTotal,
and the results reveal that custom-made ads are much more likely to promote malware than
the inherent and pop-up ads: $51.11\%$ ($23/45$) of the custom-made ads promote malware while only $34.78\%$ ($24/69$) of pop-up ads and $42.67\%$ ($32/75$) inherent ads promote malware.
\textit{This finding demonstrates the necessity of analyzing custom-made ads rather than focusing solely on ads from ad libraries.}
}

\noindent\textbf{Motivation for UI Exploration. }
\green{For ad-library ads (i.e., inherent and pop-up ads), we aim to explore whether the same ad libraries exhibit different ad promotion behaviors across apps (i.e., different apps with the same ad libraries promoting different apps).
We characterize the diversity of the promoted apps based on three aspects: functionality (app category in the app market), maliciousness (whether the app is malware), and popularity (whether the download number exceeds 1 million).
We first conduct a Pearson's chi-square test~\cite{pearson1900x} to explore the association between the use of specific ad libraries and the diversity of promoted apps, 
Specifically, we employ a lightweight static analysis~\cite{ma2016libradar} to detect the ad libraries in an app. We find that the most frequently used ad libraries in our dataset are AdMob~\cite{googleAds}, AppLovin~\cite{applovin_adformats}, and Facebook Ads~\cite{facebook_adformats}.  For each advertiser app utilizing these libraries, we randomly sampled $10$ (or all, if fewer than $10$) apps that they promoted.
The statistics reveal no significant association between the ad libraries used in the advertiser apps and the app category, maliciousness, or popularity of the promoted apps. In contrast, we discovered a significant association between the categories of advertiser apps and those of the promoted apps with a p-value of $6.02e^{-05}$.  
These findings lead us to conclude that the promotion behaviors of advertiser apps are influenced more by the individual advertiser apps than by the specific ad libraries used, aligning with findings from existing studies~\cite{book2015empirical}. This motivates us to develop UI exploration techniques for individual apps, regardless of the ad libraries used.}

\section{App Promotion Graph Construction}
In this section, we present the ad-oriented UI exploration technique in detail and describe how to construct the app promotion graph.

\subsection{Ad-Oriented UI Exploration}
\label{sec:ad_oriented_ui_exploration}
Based on the preliminary study, \sys employs a dynamic UI exploration to collect app promotion ads for building an app promotion graph rather than relying on static analysis on ad libraries.
In particular, based on the characteristics of app promotion ads, we have observed three unique challenges:
\begin{itemize}[noitemsep, topsep=1pt, partopsep=1pt, listparindent=\parindent, leftmargin=*]
\item \textit{Mobile GUI Navigation}: Identifying ad-related UI is challenging due to ads often appearing after navigating through multiple UI pages, especially when analyzing a large number of apps~\cite{guitest1,guitest2,guitest3}. 
Thus, the technique needs to effectively and efficiently navigate to the ad-related UI.
\item \textit{App Promotion Ad Detection}: App promotion ads appear in various formats, each with distinct UI layout characteristics. 
Thus, we need to design a precise mechanism in detecting app promotion ads when they show up in the UIs. 
\item \textit{Dynamic Ads Capture}: 
Ads' content is generated at runtime and continuously changes over time. 
For example, relaunching the same app after a certain period may display a different banner in the same location, promoting a different app.
Thus, UI exploration needs to repetitively visit the same UI pages to collect more ads.
\end{itemize}
We next describe how our ad-oriented UI exploration technique addresses these challenges.

\noindent\textbf{Mobile GUI Navigation.}
While previous studies focused only on ads displayed on the main page UI~\cite{liu2020maddroid}, our preliminary study on the AndroZoo dataset reveals that over $28\%$ of ads appear in UIs other than the main page. User interaction data from the Rico dataset also indicates that, on average, a user must engage in $8$ interactions to reach a UI containing ads . 

To scale up the exploration to a large number of apps while still maintaining high coverage of ads, we adopt a random exploration method combined with a depth-first exhaustive search strategy.  This strategy is shown in previous studies to be more effective in finding deeper exploration paths that are more likely to identify more ads~\cite{guitest1,paladin,monkey}.
Other more complex strategies such as program analysis techniques~\cite{gao2018android,mirzaei2012testing} or UI model refinement~\cite{guitest1,guitest2,guitest3} require heavier weight computation during UI exploration and cannot easily scale to a large number of apps.
Specifically, our technique initiates a recursive search of all UIs, starting from the main UI upon app launch. It interacts with all widgets (e.g., touch, scroll, and select). If interaction with a widget leads to navigation outside the app (excluding Google Play), our technique returns to the previous activity to explore another unexplored widget. If navigation occurs within the same app, it continues interacting with widgets in the current activity. 
This process continues until all widgets in the app's UIs have been interacted with or until a preset timeout is reached.
Following the existing practices, we also pre-register several accounts and write scripts to bypass login~\cite{li2017droidbot},
since many apps cannot navigate to the main page without passing the login.

\noindent\textbf{App Promotion Ads Detection}.
A direct approach to detecting app promotion ads is to search for ad-related keywords in the UI hierarchies of explored UIs. 
However, our analysis of the AndroZoo dataset reveals that the naming practice of ad-related widgets actually varies across different apps and developers, lacking a consistent pattern. 
Additionally, we find that app promotion ads are displayed using more than $15$ UI widgets types (\eg the most common ones are View, TextView, Button).
Thus, existing UI exploration techniques~\cite{liu2020maddroid,nath2015madscope,rastogi2016these} that focus on a small set of UI widgets such as \texttt{WebView} and \texttt{ImageView} fail to detect all the app promotion ads.

To address this challenge, we observe that all ads have the same UI design goal: encouraging users to tap on the ads to download promoted apps. 
For example, the usage of the green ``install'' button of the banner ad shown in Fig.~\ref{fig:banner}, and the orange ``INSTALL NOW!'' button for the intersititial ad shown in Fig.~\ref{fig:interstitial}.
Building on these observations, we construct a list of ad-specific keywords containing those encouraging words we encountered most frequently during the empirical study. 
Based on this list, our technique prioritizes the exploration of the widgets whose attributes (e.g., (\texttt{text}, \texttt{resource-id}, and \texttt{class})) contain any of these keywords.

After interacting with a widget, users are generally redirected to Google Play or third-party websites for app downloads. Note that Google Play does not provide direct APK downloads, and some promoted apps may be unavailable in the market due to local policy violations. 
To address this, we collect the redirection link and extract the package name of the promoted app from the link. 
For third-party websites, we manually open the link and record the package name of the promoted app. 

\noindent\textbf{Dynamic Ads Capturing}.
We observe that when an app is restarted, its app promotion ads may change (i.e., they promote a different app).
To investigate how the ads are dynamically altered, we keep restarting an app to refresh its ads.
We observe that, within a specific time period, the total number of unique ads obtained from an app remains constant~\cite{vallina2012breaking}. This suggests that ad libraries infrequently update their recommendation lists, instead maintaining them for a period of time. 

To address this challenge,  we run our exploration technique on each app within the predefined time limit ($5$ minutes in our experiment), and iteratively restart the app to repeat the detection.
We record the promoted apps from the ads until no new promoted apps are identified after a pre-defined number of iterations ($3$ in our experiment) or the maximum number of iterations is reached ($20$ in our experiment). 
We adopt this setting because it can effectively capture most of the ads recommended by the ad library within a time range.

\begin{table}[t]
\caption{Attributes from different sources}
\label{tab: features}
\resizebox{\linewidth}{!}{%
\begin{tabular}{@{}lll@{}}
\toprule
\multicolumn{1}{c}{\textbf{Source}}  & \multicolumn{1}{c}{\textbf{Name}} & \multicolumn{1}{c}{\textbf{Brief Description}}        \\ \midrule
\multirow{8}{*}{App Market} & App Name                 & The name of the app                          \\
                            & Developer Name           & The name of the developer                    \\
                            & Reviews                  & Count of user reviews                        \\
                            & Downloads                & Count of downloads                           \\
                            & Star                     & Average star rating                          \\
                            & Description              & Developer-provided app description           \\
                            & Rating                   & Age-based content rating (e.g., Teen, $18$+) \\
                            & Category                 & App category (e.g., Social, Tools)           \\ \midrule
\multirow{3}{*}{VirusTotal} & Flags                    & Count of malware flags                       \\
                            & Report                   & Results of security engine analyses          \\
                            & URL                      & URLs from the ``Interesting String'' field   \\ \midrule
\multirow{3}{*}{Code}       & Manifest                 & Content of the AndroidManifest.xml file      \\
                            & API Calls                & API calls extracted from smali code          \\
                            & Signature                & Hash of the app's certificate signature      \\ \bottomrule
\end{tabular}
}
\end{table}

\subsection{Graph Construction}

 To construct an app promotion graph,  we apply our ad-oriented UI exploration technique on the seed dataset detailed in Section~\ref{sec:dataset}. 
 Specifically, for each promoted app, we install the app based on the recorded package name from the AndroZoo dataset. We then run the exploration technique on the app to collect more app promotion ads.
This process repeats until no new apps are found in the app promotion ads.
With the collected app promotion ads, we are able to construct an app promotion graph to map out the promotion relationships within the ecosystem.

To facilitate a more in-depth analysis, we enrich the graph by collecting additional data about each app as attributes of the graph's nodes. This additional data is sourced from three distinct origins, as detailed in Table~\ref{tab: features}. 

\noindent\textbf{App Market Attributes}.
To investigate the relationship between app promotion and various factors such as organization, popularity, and functionality, we crawl the Google Play Store pages of each app and extract information such as the name of the app, the number of total downloads, and the app category.

\noindent\textbf{VirusTotal Attributes}.
To construct the groundtruth of malware and PUAs in the app promotion graph, we query VirusTotal for the security flags of each app and obtain the corresponding vendor reports. 
VirusTotal requires the APK file or its corresponding hash to perform an analysis. 
Hence, we crawl the relevant SHAs from AndroZoo. 
Notably, multiple SHAs may be associated with the same package name. 
To ensure a conservative analysis, we select the five most recent SHAs (if available) and choose the SHA with the highest number of VirusTotal flags for our analysis.

\noindent\textbf{Code Attributes}.
The ad libraries have unique code-level characteristics for their special system and network behaviors ~\cite{shao2018understanding,liu2020maddroid}. 
Hence, we obtain the API calls from the smali code of each app.
Furthermore, we download the APK file of the promoted app from AndroZoo and decompile it using Androguard~\cite{androguard}.
We next extract component names from the decompiled manifest file. 
Component names reveal reused code, such as that from the same ad library or developer.
We also extract the signature, which conveys the app's organizational information.

\begin{figure*}[h]
    \centering
    \includegraphics[width=0.85\linewidth]{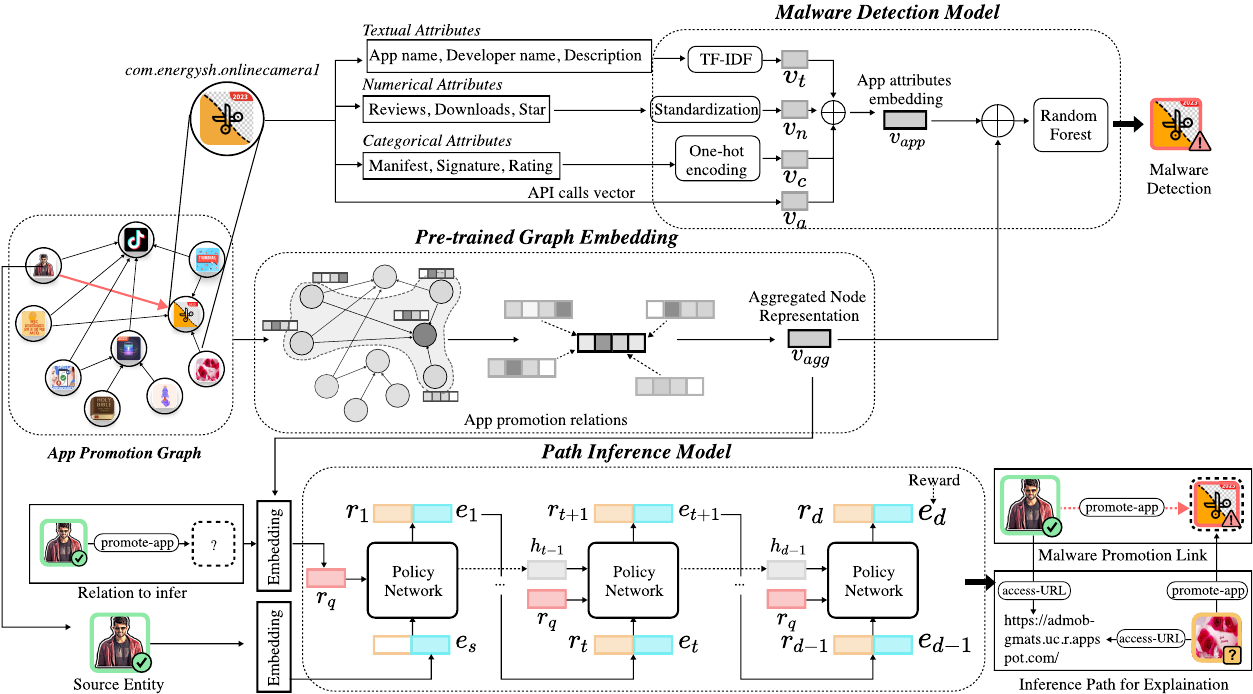}
    \caption{Overview of malware detection and promotion reasoning}
    \label{fig:overall_approach}
\end{figure*}

\section{Malware Detection and App Promotion Reasoning}
\label{sec:gnn}

As shown in Fig. \ref{fig:overall_approach}, we introduce a pre-trained graph embedding to represent each node by aggregating its neighborhood information, i.e., app promotion relations. The pre-trained node embedding is then used to perform two downstream tasks: \textit{malware detection} and \textit{app promotion reasoning}.

\subsection{Pre-trained Graph Embedding}
For a given node $a$ in the app promotion graph, we categorize the apps' attributes into three groups based on Table~\ref{tab: features}: textual, numerical, and categorical attributes. 
\begin{itemize}[noitemsep, topsep=1pt, partopsep=1pt, listparindent=\parindent, leftmargin=*]
    \item \textit{Textual attributes}: Textual attributes describe the developer, the functionality, and the purposes of an app, including the attributes of ``App Name'', ``Developer Name'', and ``Description''. To encode these attributes, we remove stopwords, apply stemming, and then employ the Term Frequency-Inverse Document Frequency (TF-IDF)~\cite{kao2007natural} technique to transform them into feature vectors, denoted as $v_t^a$.
    \item \textit{Numerical attributes}: Numerical attributes represent an app's popularity, including the attributes of ``Reviews'', ``Downloads'', and ``Star''. 
    We standardize them to form the feature vector, denoted as $v_n^a$.
    \item \textit{Categorical attributes}: Categorical attributes contain the attribute of ``Manifest'', including the activities, providers, services, receivers, permissions, and the attributes of ``Rating'', and ``Signature''. We use one-hot encoding to represent them, denoted as $v_c^a$. 
\end{itemize}
We also encode the attribute of ``API Calls" as $v_a^a$ by leveraging the technique of MaMaDroid~\cite{onwuzurike2019mamadroid}. 
The feature vectors of these four app attributes are then concatenated as the app attribute embedding $v_{app}^a = CONCAT(v_t^a, v_n^a, v_c^a, v_a^a)$. 

We employ GraphSAGE~\cite{hamilton2017inductive} to aggregate app promotion relations. Formally, for each node $a$, its one-hop neighbors are denoted by $N(a)$. We apply a mean aggregator, which takes the element-wise mean of the app attribute embedding ${v_{app}^u, u \in N(a)}$, to generate the aggregated node embedding $v_{agg}^n$, which is defined as follows: 
\begin{align}
    v_{agg}^{N(a)} = ReLU ( W_1 \cdot MEAN(v_{app}^u, u \in N(n_i)))\\
    v_{agg}^a = ReLU ( W_2 \cdot CONCAT(v_{app}^a, v_{agg}^N(a))  
\end{align}
where $W_1$ and $W_2$ are the learnable parameters of GraphSAGE.
We compute the $v_{agg}^a$ for each node as the pre-trained graph embedding that represents the semantics not only from the app attributes but also the app promotion relations, which enables the downstream tasks on the app promotion graphs.

\subsection{Malware Detection}
\label{sec:malwaredetection}

Malware detection refers to detecting whether a promoted app is PUA/malware. 
Conventional solutions in the industry (e.g., VirusTotal) rely on signature analysis.
Consequently, they often fail when the apps in question are freshly developed or recently updated, which is common for apps promoted by ads. 
Moreover, existing academic Android malware detection studies~\cite{hou2017hindroid,onwuzurike2019mamadroid,xu2019droidevolver,zhang2020enhancing} do not consider app promotion relations, making them less effective in detecting malware through app promotions.

To address this challenge, we leverage both app attributes and app promotion relations from our app promotion graph to develop a comprehensive malware detection model.
Specifically, we concatenate the app attribute embedding $v_{app}$ and the pre-trained node embedding $v_{agg}$ to form a unified feature vector for each app. 
This feature vector captures both the app attributes and the promotion relations.
We further concatenate the united feature vectors for both the advertiser app and the promoted app as the final vector to enhance the app promotion information. 
Due to the nature of the app promotion ecosystem, our input data are imbalanced, and benign apps take a majority of them. 
To mitigate this issue, we used Random Forest~\cite{ho1995random} as our final classifier. 
As an ensemble method, Random Forest combines the results of multiple decision trees, thereby reducing noise and overfitting caused by the imbalanced data.


\subsection{App Promotion Reasoning}
\label{sec: Knowledge_Graph_Construction}
The app promotion graph contains rich information on app attributes and app promotion relations.
To reveal how such information contributes to promoting specific apps, especially malware, \sys transforms the app promotion graph into a promotion inference graph (\ig) and establishes a path inference model to infer malware promotion mechanisms.
Particularly, this path inference model works together with our malware detection model to enhance \sys's capability in analyzing app promotion ads, and we summarize the benefits of our path inference model as the following two aspects: 
\begin{itemize}[noitemsep, topsep=1pt, partopsep=1pt, listparindent=\parindent, leftmargin=*]
\item \textit{Enhancing Explainability in Malware Detection}: existing AI techniques for malware detection suffer from a lack of  
explainability~\cite{gerlings2020reviewing}. Our path inference model addresses this gap by inferring malware promotion mechanisms (e.g., ad libraries, shared developers). The observed recurring mechanism increases the confidence of our malware detection.
\item \textit{Predicting Missing App Promotion Links}: Without access to the complete list on the ad server, our UI exploration technique is limited in capturing all app promotion links. Moreover, even when such links are identified, the process can be time-consuming. Our path inference model steps in to predict these missing links, streamlining the construction of the app promotion graph. 
Inferring unseen malware promotion paths also aids in identifying malware with similar promotion mechanisms.
\end{itemize}



\begin{table}[t]
\centering
\caption{Promotion Inference Graph (\ig) relations}
\label{tab:kg_relations}
\resizebox{\linewidth}{!}{%
\begin{tabular}{@{}lp{5cm}@{}}
\toprule
\multicolumn{1}{c}{\textbf{Relation}} & \multicolumn{1}{c}{\textbf{Definition}}                                                          \\ \midrule
\textit{R1: promote-app}                       & An app promotes another via app promotion ads                                                         \\
\textit{R2: has-sig}                           & An app has a digital signature, verifying its authenticity and integrity             \\
\textit{R3: has-manifest}                      & An app has a manifest file containing activities, providers, receivers, services, and permissions \\
\textit{R4: access-URL}                        & An app has access to a URL                                                                                \\
\textit{R5: isin-category}                     & An app belongs to an app category in Google Play                                                  \\
\textit{R6: VT-flag}                           & A security engine flags an app as malware
\\
\textit{R7: create-app}                        & A developer creates an app                                                                                \\
\textit{R8: develop-category}                  & A developer creates an app belonging to a category                               \\
\textit{R9: use-URL}                           & A developer creates an app which accesses a URL                                                \\ \bottomrule
\end{tabular}
}
\end{table}

\noindent\textbf{\ig Construction.}
We build the \ig by extracting the app attributes as entities and adding the relations as denoted in Table~\ref{tab:kg_relations}.
Formally, we define \ig $ = (E, R)$, where $E$ is the set of entities and $R$ is the set of relations.

\noindent\textbf{Path Inference}
We train a path inference model to infer the malware promotion mechanisms and predict missing links over \ig, as shown in Fig.~\ref{fig:overall_approach}. Following the model structure of Multi-Hop~\cite{lin2018multi}, our model employs a reinforcement learning algorithm~\cite{williams1992simple}. This is designed to infer the destination entity $e_d \in E$ based on the input source entity $e_s \in E$ and the relation $r_q \in R$ we desire to infer. At each time point $t$, the model decides an entity $e_t \in E$ and the relation $r_t \in R$ that leads to $e_t$. Eventually, the model outputs an inference path ($e_s, r_1, e_1, ..., r_t, e_t, ..., r_d, e_d$) to connect $e_s$ and $e_d$, where the relation between $e_s$ and $e_d$ aligns with $r_q$.

We use word2vec~\cite{mikolov2013efficient} to convert the indices of the source entities and relations into a vector form. We then augment the entity representation by concatenating the vectors of app entities with our pre-trained graph embedding $v_{agg}$. For other entities and relations without $v_{agg}$, we append vectors of zeros to maintain a uniform representation.

During the training time, we employ a policy network~\cite{williams1992simple} to output entities and relations for each time point $t$. 
Specifically, the input of the policy network consists of the pair of relation and entity ($e_t$, $r_t$) and the hidden state $h_{t-1}$, which represents the search history ($e_s, r_1, e_1, ..., r_{t-1}, e_{t-1}$). 
Given the desired relation $r_q$, the policy network then samples the pair of relation and entity ($e_{t+1}$, $r_{t+1}$) for the next time point and outputs the search history $h_t$. 
This process repeats until it reaches the destination entity $e_d$. 
To ensure accurate path-finding, the policy network gets a reward of $1$ for observed triplets ($e_s$, $r_q$, $e_d$) in the \ig. Otherwise, it gets an approximated soft reward~\cite{lin2018multi}. 
Soft rewards account for the app promotion graph's dynamic and incomplete nature, enabling the model to identify unseen paths in training, enabling the model to predict missing links.
Eventually, the traversed relations and entities constitute the inference path ($e_s, r_1, e_1, ..., r_t, e_t, ..., r_d, e_d$), explaining the potential reason for app promotion.

During the inference time, we input pairs of source entities and relations. The path inference model encodes them through the policy network, then performs a breadth-first search to decode and find destination entities and generates the inference paths. If the output entities contain malware, we review the inference path and identify the promotion mechanisms.
To diversify destination entities, we apply a post-processing technique~\cite{lin2018multi} that computes a list of unique destination entities and assigns the maximum score among all paths leading to each unique entity. The final outputs are the top-ranked unique destination entities with the inference path.

\section{Evaluation Setup}

\subsection{Research Questions}
We seek to evaluate the effectiveness of \sys in constructing the app promotion graph and detecting the ad-promoted malware.
Specifically, we aim to answer the following research questions:

\begin{itemize}[noitemsep, topsep=1pt, partopsep=1pt, listparindent=\parindent, leftmargin=*]
    \item \textbf{RQ1:} How effectively can \sys detect malware promoted by ads? 
    \item \textbf{RQ2:} How apps, especially malware, are promoted in the constructed app promotion graph?
    \item \textbf{RQ3:} How the apps with app promotion ads found by \sys evolve across time? Among these apps, are there any new apps that have not been scanned by VirustTotal? 
    \item \textbf{RQ4:} How effectively can \sys build the app promotion graph?
    \item \textbf{RQ5:} How effectively can \sys's path inference model predict promotion links? Can these predicted links be used to reason about how apps are promoted through ads?
\end{itemize}
We next present our dataset and implementation, and then describe our evaluation findings in detail.


\subsection{Dataset}
\label{sec:dataset}
\noindent\textbf{Seed Dataset}.
We collect apps from AndroZoo as our study subjects since AndroZoo is well-maintained and regularly updated with various versions of apps. 
Within our affordable effort, we target the apps released from January 1st, 2018 to February 3rd, 2023. 
Following the practice of the existing works~\cite{ishii2017understanding,liu2020maddroid}, we curate the dataset for three app classes based on the malware flags from VirusTotal: (1) \textit{Malware}, which is flagged by at least $10$ engines; (2) \textit{PUAs}, which are flagged by $1$-$9$ engines; and (3) \textit{Benign Apps}, which are not flagged by any engine.
By distinguishing between  ``PUA'' and ``malware'', we can analyze how potentially harmful apps are exploited to promote those with actual malicious behaviors, gaining more understanding of the app promotion ecosystem. 
In total, we construct a seed dataset consisting of $36,000$ apps, with $12,000$ for each app class.


\noindent\textbf{App Promotion Graph Dataset}.
Table~\ref{tab:dataset_summary} shows a summary of our dataset.
Out of the $36,000$ seed apps, we successfully executed $15,344$ apps on our devices.
As some apps were released as early as 2018, they are not properly upgraded for the recent Android versions and cannot be executed.
Most of the other apps failed the executions as they crashed during the launch.
In the executed apps, our UI exploration technique identifies $2,420$ apps that have app promotion ads and further collects $3,859$ additional apps promoted by them. 
Among these, there are $271$ apps that not only have app promotion ads but are also promoted by other apps.


In total, our constructed app promotion graph consists of $6,008$ nodes/apps, including $464$ malware, $1,042$ PUAs, and $3,961$ benign apps.
Note that the total number of apps ($6,008$) is not a sum of the number of apps ($5,467$) in the malware, PUA, and benign classes. 
This discrepancy is due to some promoted apps ($541$) not being archived in AndroZoo at the time of our analysis, which we will further elaborate in Section~\ref{sec:temporal_analyais}.
From these apps, our technique collects $18,627$ instances of app promotion ads.
In particular, $22.2\%$ of the ads are linked to malware ($520$) and PUAs ($3,616$), while the remaining ads are tied to benign apps ($13,054$). 
Additionally, our technique records screenshots, timestamps, and interaction traces during app exploration to confirm and evaluate the collected app promotion ads.

\noindent\textbf{Malware Detection Dataset}.
Our malware detection model is trained and tested on the app promotion graph with $5,467$ nodes, i.e., the number of nodes that have VirusTotal labels, and $18,627$ links, using stratified 10-fold cross-validation.  The path inference model, based on a graph with $127,100$ entities and $565,739$ relations, is split in an $80$:$10$:$10$ ratio for training, validation, and testing.

\subsection{Implementation}
We implement the ad-oriented UI exploration technique upon DroidBot~\cite{li2017droidbot}, which has been well-maintained up to November 2023. DroidBot is compatible with all Android API versions and offers an interface for implementing customized exploration strategies.
Since ad libraries employ emulator detection mechanisms to prevent fraudulent ad traffic, showing only test ads on emulators, we conduct our app exploration on real devices. 
Specifically, we use six Samsung Galaxy A13 smartphones running Android 11 for our research, which extends over a period of more than three months.
We use mitmproxy to record the redirection links after interacting with UI widgets~\cite{mitmproxy}.
We leverage networkx~\cite{hagberg2008exploring} to build the app promotion graph.

To construct the pre-trained graph embedding, we use two GraphSAGE~\cite{hamilton2017inductive} convolution layers with $128$-dimension embedding. The dropout rate for each layer is $0.5$. We apply the Adam optimizer~\cite{kingma2014adam} with learning rate of $0.01$ and weight decay of $0.0005$. 
We follow the model architecture in Multi-Hop~\cite{lin2018multi} to implement the path inference model. Each edge is regarded as bidirectional. We only keep the top-$256$ neighbors for each entity with the highest PageRank~\cite{page1998pagerank} scores to prevent GPU memory overflow. 
The embedding size of all entities and relations is $200$. Xavier initialization~\cite{glorot2010understanding} is used for initializing all neural network layers within the model. The SOTA embedding model, DistMult~\cite{DistMult}, is employed to attain the soft reward. We apply a three-layer LSTM to encode hidden states, each with a dimension of $200$. We apply the Adam optimizer with a learning rate of $0.001$ and a mini-batch size of $32$. The dropout rate for the neural layers is set at $0.1$. The decoding breadth-first search size is $128$. 

\begin{table}[t]
\caption{Dataset summary}
\large
\label{tab:dataset_summary}
\resizebox{\linewidth}{!}{
\begin{tabular}{@{}lrlr@{}}
\toprule
\# Seed Apps                    & $36,000$ & \# Apps Executed                & $15,344$ \\ \midrule
\# Apps Having App Promotion Ads              & $2,420$  & \# Apps Being Promoted          & $3,859$  \\ \midrule
\# Nodes of App Promotion Graph & $6,008$  & \# Links of App Promotion Graph & $18,627$ \\ \bottomrule
\end{tabular}
}
\end{table}

\begin{table}[t]
    \centering
    \caption{Comparison of malware detection baselines and ablation study. The $-$promotion denotes \sys trained without app promotion relations. The $\rightarrow$DGI, $\rightarrow$GRACE, and $\rightarrow$MVGRL denote \sys with the embedding method, GraphSage, replaced with DGI, GRACE, and MVGRL, respectively}
    \large
    \resizebox{\linewidth}{!}{
\begin{tabular}{clllll}
\toprule
                                                                             & \multicolumn{1}{c}{\textbf{Approaches}} & \multicolumn{1}{c}{\textbf{Accuracy}} & \multicolumn{1}{c}{\textbf{Precision}} & \multicolumn{1}{c}{\textbf{Recall}} & \multicolumn{1}{c}{\textbf{F1 score}} \\
\midrule
\multirow{7}{*}{Baselines} & Symantec                                     & $96.99$                     & $81.66  $          & $69.01   $          & $74.80 $           \\
                                   & Lionic                                       & $96.72   $                  & $74.64$            & $74.64  $           & $74.64 $           \\
                                   & McAfee                                       & $95.99      $               & $69.56$            & $67.60 $            & $68.57    $        \\
                                   & Avira                                        & $94.26     $                & $53.57$            & $84.50   $          & $65.57  $          \\
                                   & K7GW                                         & $93.63$                     & $50.41$            & $85.91$             & $63.54 $           \\
                                   & DroidEvolver~\cite{xu2019droidevolver}                                 & $75.48_{\pm 7.12}$                     & 
                                   $72.92_{\pm7.96}$  & $70.93_{\pm 11.39}$ & $71.21_{\pm 6.96}$ \\
                                   & MaMaDroid~\cite{onwuzurike2019mamadroid}                                    & $79.38_{\pm 7.33}$                     & 
                                   $75.48_{\pm6.32}$  & $78.41_{\pm 9.54}$ & $76.58_{\pm 6.14}$ \\
                                   & \blue{ANDRUSPEX~\cite{shen2021andruspex}}                                    & \blue{$95.15_{\pm 1.24}$}                     & 
                                   \blue{$95.32_{\pm1.14}$}  & \blue{$88.79_{\pm 3.19}$} & \blue{$92.48_{\pm 3.19}$} \\
                                   
\midrule
\multirow{5}{*}{\makecell[l]{Ablation\\ Study}}  & $-$ promotion &   $96.29_{\pm 1.07}$        & $95.27_{\pm 3.68}$&  $86.01_{\pm 7.23}$  & $90.14_{\pm 7.23}$ \\
                                   & $\rightarrow$DGI~\cite{dgi}                   & $97.47_{\pm 0.61}$          & $99.10_{\pm 2.19}$ & $91.43_{\pm 6.82}$  & $94.96_{\pm 6.82}$ \\
                                   & $\rightarrow$GRACE~\cite{mvgrl}               & $97.45_{\pm 0.66}$          & $\text{99.82}_{\pm 0.55}$ & $91.43_{\pm 6.64}$  & $95.30_{\pm 6.64}$ \\
                                   & $\rightarrow$MVGRL~\cite{dasgo2018ICLR}                  & $97.38_{\pm 0.65}$          & $98.57_{\pm 2.48}$ & $90.90_{\pm 7.27}$  & $94.40_{\pm 7.27}$ \\
                                   & \textbf{\sys}                                & $\textbf{97.74}_{\pm 0.62}$ & $99.44_{\pm 1.67}$ & $\textbf{91.78}_{\pm 7.02}$  & $\textbf{95.31}_{\pm 7.02}$ \\
\bottomrule
\end{tabular}
}
\label{tab:malware_detection}
\end{table}

\section{Results}
\subsection{RQ1: Malware Detection}




\label{sec:eval_malware_prediction}
\noindent\textbf{Comparison to Baselines}.
We compare \sys's performance against the five security engines of VirusTotal with the best F1 scores. 
\blue{Additionally, we compare \sys with three state-of-the-art (SOTA) malware detection approaches: MaMaDroid~\cite{onwuzurike2019mamadroid}, DroidEvolver~\cite{xu2019droidevolver} and ANDRUSPEX~\cite{shen2021andruspex}. 
For a fair comparison, we re-implement these approaches using their open-source code and employ Random Forest classifiers, aligning with their original configurations. }

As illustrated in Table~\ref{tab:malware_detection}, \sys outperforms all the compared baselines, achieving a $97.74\%$ accuracy, a $99.44\%$ precision, a $91.78\%$ recall, and a $95.31\%$ F1 score.
In contrast, though the five security engines achieve high accuracy, all of the baselines suffer from low precision, recall, and F1 score values, suggesting a poor performance in malware detection.
The superior performance of \sys can be attributed to ``App Market'' attributes, which are among the most important features of its random forest classifier. 
For example, some malware is excluded from the Google Play store or exhibits distinctive features such as low star ratings.
Overall, these findings demonstrate the superiority of \sys over commercial security products and SOTA malware detection approaches.



\noindent\textbf{Ablation Study}.
We conduct an ablation study to measure the importance of using app promotion relations and compare the effectiveness of different embedding methods for the promotion relations, i.e., DGI~\cite{dgi}, GRACE~\cite{grace}, MVGRL~\cite{mvgrl}.
As shown in  Table~\ref{tab:malware_detection}, using app promotion relations helps detect extra malware samples, as it significantly increases the recall of \sys from $.01\%$ to $91.78\%$, and F1 score from $90.14\%$ to $95.31\%$.
Furthermore, while all of them significantly exceed the performance of \sys trained without app promotion relations, GraphSage, adopted by our approach, outperforms other embedding methods, 
\textit{These results demonstrate that app promotion relation is an indispensable feature in detecting ad-promoted malware.}

\begin{figure}[t]
    \centering
    \includegraphics[width=0.9\columnwidth]{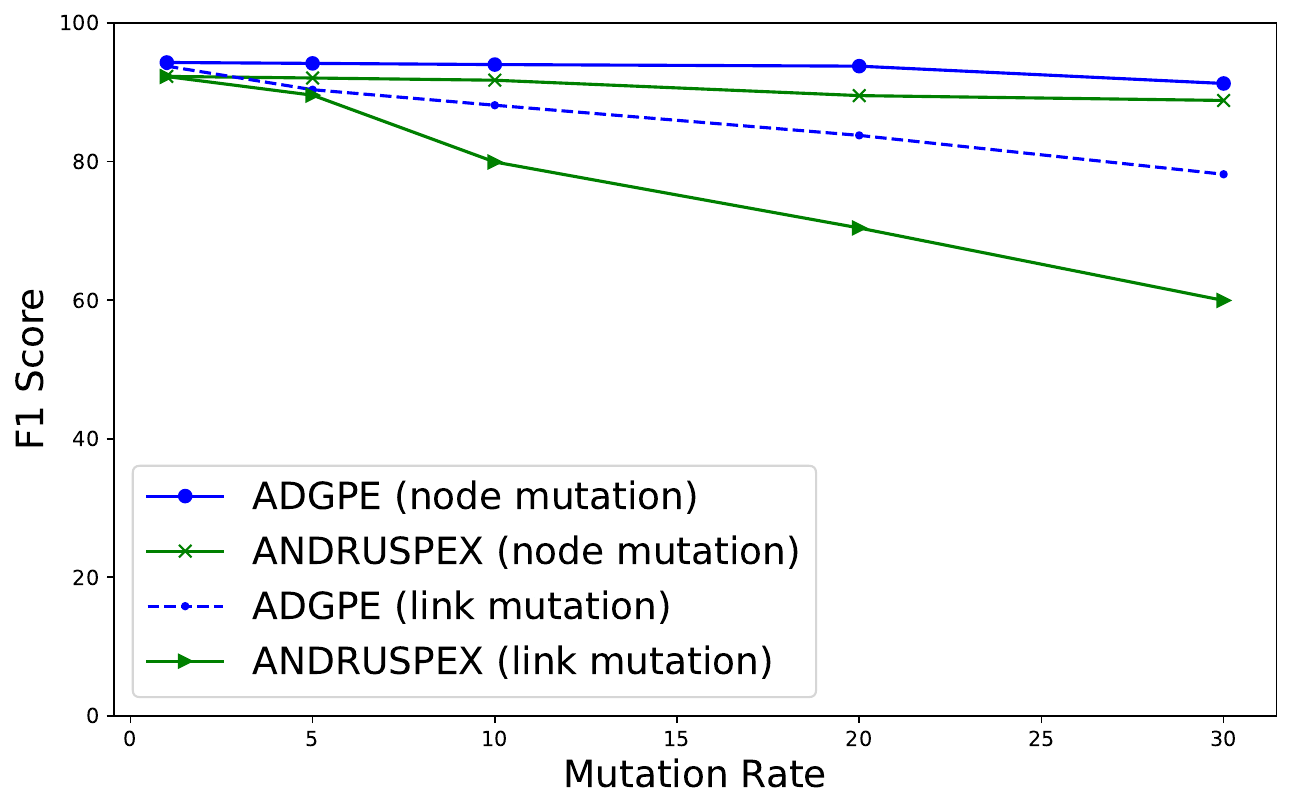}
    \caption{\blue{Robustness comparison for \sys and ANDRUSPEX}}
    \label{fig:mutation}
\end{figure}

\noindent\textbf{Robustness Analysis.}
\blue{To evade detection, attackers can manipulate the node attributes (e.g., by changing app descriptions of malware) or links (e.g., by letting malware promote benign apps).
We therefore evaluate the robustness of  \sys by mutating nodes and links. Specifically, we apply Gaussian noise to node attributes and perform random link swapping based on the out-degree of each node at rates of $1\%$, $5\%$, $10\%$, $20\%$, and $30\%$. 
As shown in Fig.~\ref{fig:mutation}, for node mutations, \sys maintains high F1 scores, with a slight drop from $94.29\%$ at $1\%$ to $91.26\%$ at $30\%$. In contrast, link mutations significantly affect performance, decreasing F1 scores from $93.76\%$ at $1\%$ to $78.16\%$ at $30\%$. This degradation is more profound in ANDRUSPEX~\cite{shen2021andruspex}, whose F1 score drops from $92.26\%$ at $1\%$ to $59.95\%$ at $30\%$, demonstrating that \sys outperforms ANDRUSPEX in adversarial settings. To defend against such attacks, we can also integrate adversarial training mechanisms by incorporating synthesized adversarial examples in the training data to mitigate the impacts of these attacks.}

\noindent\textbf{Case Study on Detected Malware}
Besides identifying malware promoted via ad libraries, \sys additionally identifies malware promoted via custom-made ads, which are overlooked by prior studies~\cite{liu2020maddroid,rastogi2016these,zhao2023mobile}.
For example, \sys identifies a magnet searcher app, available on Google Play~\cite{magnet_searcher}, as malware.
Noteworthy, this app itself does not exhibit any malicious behavior. However, clicking on any search result within the app redirects users to a third-party website where they are prompted to download a malicious downloader~\cite{malicious_downloader_magnet_searcher}.    
This suggests that the magnet searcher app, which seems harmless on the surface, is actually a malware dropper. This strategy allows it to bypass traditional malware detection models. 
By using the app promotion graph, \sys is able to detect these hidden threats effectively.

\begin{table}[t]
\large
\caption{Statistical analysis of app promotions by hop distance and app class.  Dst/Src represents the destination/source of an app promotion}
\label{tab:promotestats}
\resizebox{\linewidth}{!}{%
\begin{tabular}{@{}llrrrrrr@{}}
\toprule
\multicolumn{1}{c}{}          & \multicolumn{1}{c}{}                            & \multicolumn{3}{c}{\textbf{\# App Promotions}}                                                                & \multicolumn{3}{c}{\textbf{Promotion Probability ($\%$)}}                                                   \\ \midrule
\multicolumn{1}{c}{\textbf{}} & \multicolumn{1}{c}{\textbf{\diagbox{Dst}{Src}}} & \multicolumn{1}{c}{\textbf{Benign}} & \multicolumn{1}{c}{\textbf{PUA}} & \multicolumn{1}{c}{\textbf{Malware}} & \multicolumn{1}{c}{\textbf{Benign}} & \multicolumn{1}{c}{\textbf{PUA}} & \multicolumn{1}{c}{\textbf{Malware}} \\ \midrule
\multirow{3}{*}{Hop=1}        & Benign                                          & $8,206$                             & $3,052$                          & $1,796$                              & $72.53$                             & $64.57$                          & $69.45$                              \\
                              & PUA                                             & $2,997$                             & $1,320$                          & $736$                                & $26.49$                             & $27.92$                          & $28.46$                              \\
                              & Malware                                         & $111$                               & $355$                            & $54$                                 & $0.98$                              & $7.51$                           & $2.09$                               \\ \midrule
\multirow{3}{*}{Hop=2}        & Benign                                          & $14,735$                            & $10,046$                         & $2,918$                              & $73.08$                             & $70.82$                          & $70.64$                              \\
                              & PUA                                             & $5,248$                             & $3,656$                          & $1,133$                              & $26.03$                             & $25.77$                          & $27.43$                              \\
                              & Malware                                         & $181$                               & $484$                            & $80$                                 & $0.90$                              & $3.41$                           & $1.9$                                \\ \midrule
\multirow{3}{*}{Hop=3}        & Benign                                          & $20,686$                            & $15,089$                         & $4,250$                              & $73.41$                             & $70.93$                          & $71.55$                              \\
                              & PUA                                             & $7,276$                             & $5,685$                          & $1,601$                              & $25.82$                             & $26.72$                          & $26.95$                              \\
                              & Malware                                         & $216$                               & $500$                            & $89$                                 & $0.77$                              & $2.35$                           & $1.50$                               \\ \midrule
\multirow{3}{*}{Hop=4}        & Benign                                          & $26,407$                            & $18,384$                         & $5,501$                              & $74.19$                             & $71.50$                          & $73.35$                              \\
                              & PUA                                             & $8,915$                             & $6,809$                          & $1,903$                              & $25.05$                             & $26.48$                          & $25.37$                              \\
                              & Malware                                         & $272$                               & $518$                            & $96$                                 & $0.76$                              & $2.01$                           & $1.28$                               \\ \midrule
\multirow{3}{*}{Hop=5}        & Benign                                          & $30,731$                            & $20,497$                         & $6,282$                              & $74.25$                             & $71.75$                          & $73.35$                              \\
                              & PUA                                             & $10,353$                            & $7,539$                          & $2,185$                              & $25.01$                             & $26.39$                          & $25.51$                              \\
                              & Malware                                         & $306$                               & $532$                            & $98$                                 & $0.74$                              & $1.86$                           & $1.14$                               \\ \bottomrule
\end{tabular}
}
\end{table}

\subsection{RQ2: Malware Promotion}
\label{sec: app_promotion_in_ads}

\noindent\textbf{Overall Statistics}.
To understand how different app classes are promoted in ads,
we conduct a statistical analysis of the constructed app promotion graph
and quantify the likelihood of an app class promoted by another app class within $k$ hops in the app promotion graph by computing the corresponding promotion probability.
For example, the promotion probability of a PUA ($\mathbb{P}$) directly promoting malware ($\mathbb{M}$) at $ \text{Hop}=1 $ is computed as 
P$(\mathbb{P}\rightarrow\mathbb{M} ) = \frac{|\mathbb{P}\rightarrow\mathbb{M}|}{|\mathbb{P}\rightarrow\mathbb{M}| + |\mathbb{P}\rightarrow\mathbb{P}| +|\mathbb{P}\rightarrow\mathbb{B}|} $,
where $|\mathbb{P}\rightarrow\mathbb{M}|$ represents the number of ads in PUAs that promote malware, $|\mathbb{P}\rightarrow\mathbb{P}|$ represents the number of ads in PUAs that promote PUAs, and $|\mathbb{P}\rightarrow\mathbb{B}|$ represents the number of ads in PUAs that promote benign apps. 
This yields a promotion probability of $ 7.51\% = \frac{355}{3052+1320+355} $. 
Note that we restrict our analysis to at most $5$ hops, as extending beyond this limit shows marginal changes in discovering ads promoting new apps. 

\noindent\textbf{Malware Promotion}.
\label{sec: mal_promote}
 As evidenced by the $Hop=1$ section of Table~\ref{tab:promotestats}, merely
P$(\mathbb{B}\rightarrow\mathbb{M})=0.98\%$ of ads within benign apps directly promote malware. 
In contrast, P$(\mathbb{B}\rightarrow\mathbb{P})=26.49\%$ benign apps promote PUAs, 
which have a substantially larger probability, P$(\mathbb{P}\rightarrow\mathbb{M})=7.51\%$, to further promote malware. 
This reveals a covert malware promotion route in the app promotion graph. 
Moreover, our dataset reveals that PUAs either directly ($Hop=1$) or indirectly ($Hop>1$) promote the vast majority of malware:  $ 90.46\% $ $ (484/535) $ within two hops and $ 99.25\% $ $ (532/535) $ within five hops. 
\textit{These findings clearly show that engaging with PUAs through app promotions significantly increases the risk of malware installation. }

To better illustrate the risk of malware promotion in ads, we estimate the probability for users to encounter malware in the Google Play market, and compare it to the probability of encountering malware from app promotion ads. 
Using a conservative calculation, we find that merely $0.002\%$ of all apps ($11,014$ out of $5,151,555$) released on the Google Play market from January 1, 2018, to February 3, 2023, are malware. Considering that Google Play downloads are predominantly from popular apps~\cite{zhong2013google}, the actual probability of encountering malware there is even lower than $0.002\%$. 
In contrast, the probability of users encountering a malware promoted by app promotion ads is significantly higher: $7.51\%$ from PUAs, $2.09\%$ from malware, and $0.98\%$ from benign apps.
While these probabilities are not very high, due to the huge user base of Google Play~\cite{gplaymalware,googleAds}, a handful of malware, such as trojan, can easily harm millions of users.
\textit{As the probability of encountering malware through ads exceeds the Google Play market rates by over $100$ times, our findings demonstrate that app promotions pose a significant security risk to the users,
and more regulations should be applied to vet the apps being promoted in ads.}

  \begin{figure}[t]
    \centering
    \begin{subfigure}[]{0.32\linewidth}
      \centering
      \includegraphics[width=\linewidth]{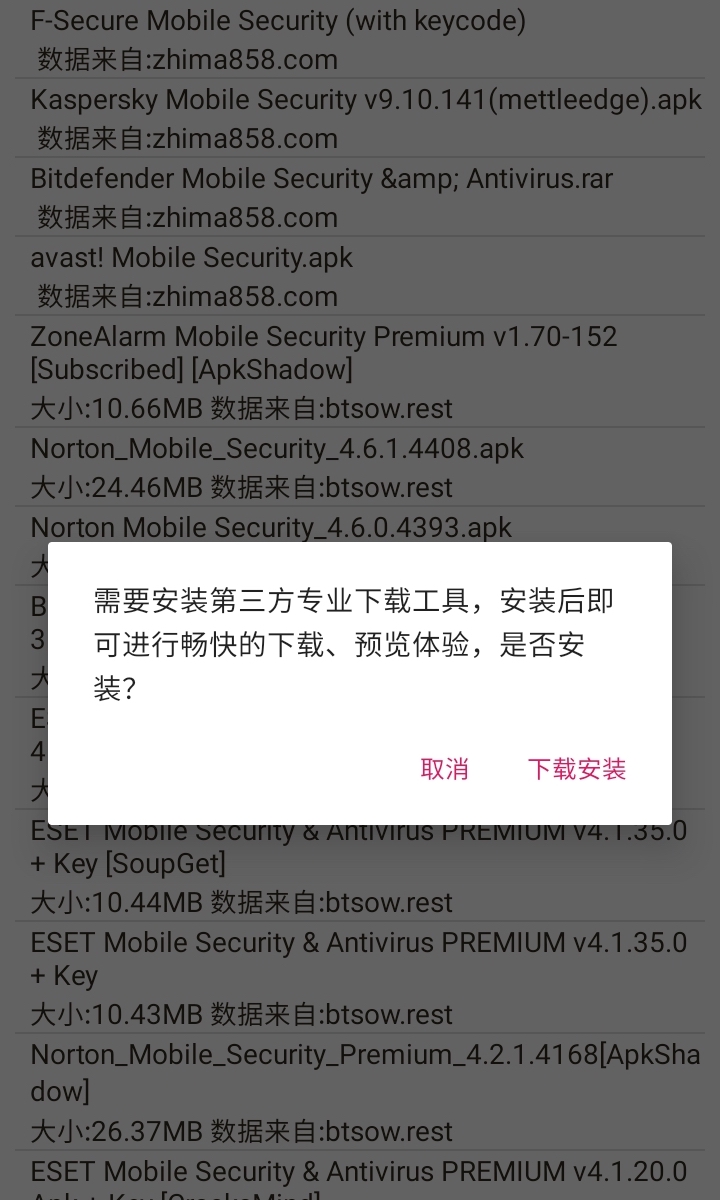}
      \caption{Trojan Dropper}
      \label{fig:magnet_customized}
    \end{subfigure}
    \begin{subfigure}[]{0.32\linewidth}
      \centering
      \includegraphics[width=\linewidth]{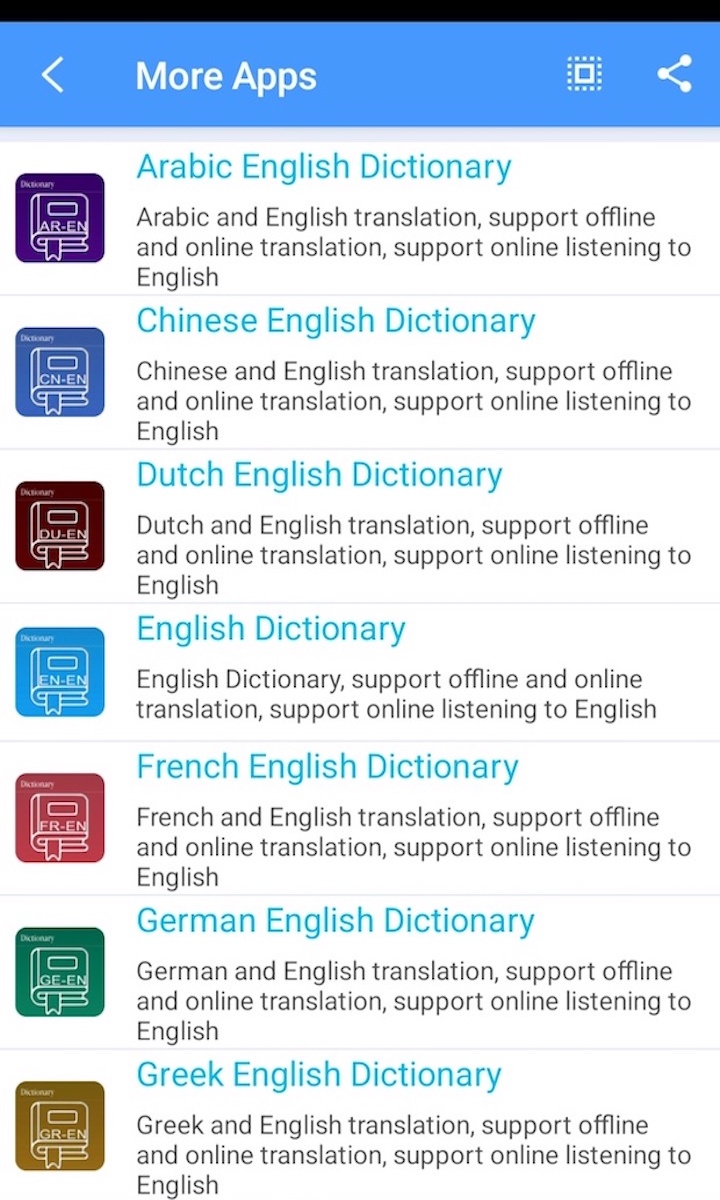}
      \caption{Trojan}
      \label{fig:lomol_appwall}
    \end{subfigure}
    \begin{subfigure}[]{0.32\linewidth}
      \centering
      \includegraphics[width=\linewidth]{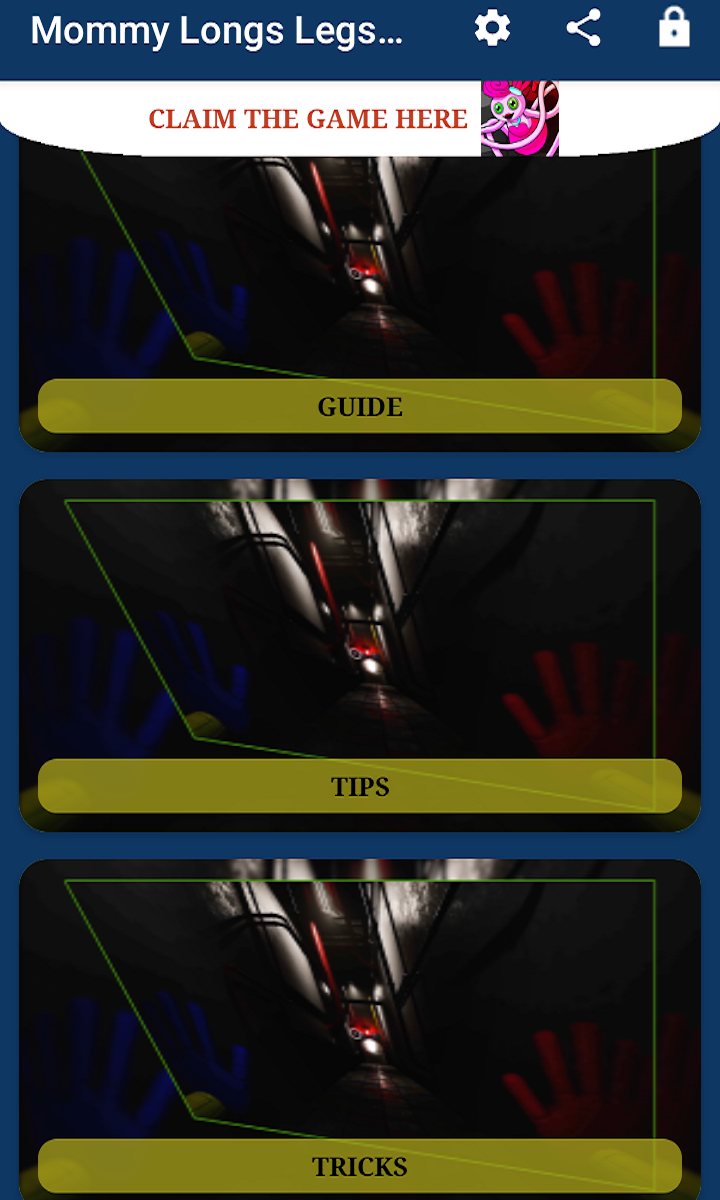}
      \caption{Aggressive Adware}
      \label{fig:wallpaper}
    \end{subfigure}
    \caption{Case study on malware promotion}
    \label{fig:case_malware}
  \end{figure}

\noindent\textbf{Case Study.}
We further conduct a case study on what are the unwanted behaviors of the promoted PUAs and malware, as well as the strategies employed in their promotions.

 \begin{itemize}[noitemsep, topsep=1pt, partopsep=1pt, listparindent=\parindent, leftmargin=*]

\item \textit{Trojan:}
More than half ($53\%$) of the PUAs/malware are labeled as trojans by VirusTotal. Interestingly, many are labeled as trojans primarily due to the presence of code obfuscation techniques, such as the use of the ``jiagu" library, rather than any malicious behavior identified based on our manual analysis. 
However, our investigation did uncover a subset of genuine trojans ($93$ in total), all from the developer ``lomol language" or ``lomol translator." 
These apps typically masquerade as dictionaries or language translators. 
For instance, one app disguises itself as a Chinese dictionary and employs code obfuscation techniques that hinder static analysis~\cite{chinese_dictionary_trojan}.
By executing the app, we observe that the app scans network information (e.g., reading files under the \texttt{/sys/class/net/directory}), and system information (e.g., executing the shell command \texttt{ls /}), and try to gain root access by executing \texttt{su}.

\noindent\textbf{Promotion Strategy:}
These ``lomol" apps adopt a unique promotion strategy by developing custom-made app promotion ads and using their own ad library (please refer to Fig.~\ref{fig:lomol_appwall} for an example) instead of employing the common ad libraries.
 
\item \textit{Aggressive Adware}: 
Approximately one-third ($32\%$) of PUAs/malware within the app promotion graph are labeled as adware by VirusTotal, primarily due to their aggressive advertising behaviors. 
Intriguingly, we observe that adware providing similar functionality such as wallpaper downloading and barcode scanning often shares common package names and UIs. 
For instance, we identify a cluster of $59$ PUAs/malware with package names incorporating the title of a popular game followed by terms like ``guide," ``trick," or ``hint" (see Fig.~\ref{fig:wallpaper}). 
Additionally, in another group of $13$ PUAs/malware, each adware provides downloads of wallpapers related to well-known movies or anime (e.g., Venom, Sonicex).

\noindent\textbf{Promotion Strategy:} The main activity names of these apps all contain the string ``AppsGeyser", a platform that advertises itself as a rapid app creation tool~\cite{appgeyser_no_code_maker}. 
This finding suggests the mass-production strategy employed by ad monetizers, leveraging streamlined app creation processes and distributing through app promotion ads to target specific user demographics, such as children and enthusiasts of games and cartoons.

\item \textit{Rogue Security Software}: 
There are five PUAs with ``fancyclean" included in their package names, all developed by ``Fancy Mobile Apps''. These apps pretend to be security engines or phone cleaners but force users to watch $30$-second video ads upon performing almost every operation (see Fig.~\ref{fig: app promotion ad example}). 

To verify whether these apps provide the promised functionality, we deploy one of the PUAs, namely ``Fancy Security \& Antivirus", on a testing phone that we deliberately install viruses, and run the app to scan the phone.
We find that the app does not actually scan the phone and simply overlooks the installed viruses.
Note that McAfee reported a similar group of ``HiddenAds'' malware in 2022~\cite{mcafee_report_rogue_security_softwares}. Interestingly, the ``fancyclean" apps in our dataset not only exhibit similar behaviors but also have similar package names, app names, and icons to the malware in the report.

\noindent\textbf{Promotion Strategy:} 
We find that the ``Fancy Mobile Apps" developer employs a centralized promotion strategy. 
The developers leverage their most popular flagship app, ``Fancy Battery: Cleaner, Secure" with a Facebook profile to attract traffic and promote other apps~\cite{fancy_battery_facebook_profile}. Although it has been removed from Google Play, cached data reveals that this flagship app receives over $13$ million downloads.

\end{itemize}

\subsection{RQ3: Temporal Analysis}
\label{sec:temporal_analyais}
We conduct a temporal analysis to understand how app promotion ads evolve over time.
Initially, $1,334$ apps were identified to promote either PUA or malware in Feb 2023. 
Note that the other part of this paper relies on data and findings from this February experiment. 
We reran the ad-oriented UI exploration technique in August 2023, updating our dataset based on the AndroZoo database, which is refreshed daily.
Overall, out of the initial $1,334$ apps, $734$ apps still have app promotion ads, while the total number of promoted apps has dropped from $3,212$ to $1,228$. The amount of ads promoting benign apps, PUAs, and malware has decreased by roughly $30\%$, $45\%$, and $61\%$, respectively.  This is mainly because the developers stopped maintaining the app promotion ads, often due to the release of newer versions or the apps being removed from Google Play.
For instance, the Chinese dictionary trojan, mentioned in RQ2, was found to promote $107$ unique apps before its update.  However, in August, the February version of the app displayed no app promotion ads while the latest version, released in August displayed such ads.

\begin{figure}[t!]
    \centering
    \includegraphics[width=\columnwidth]{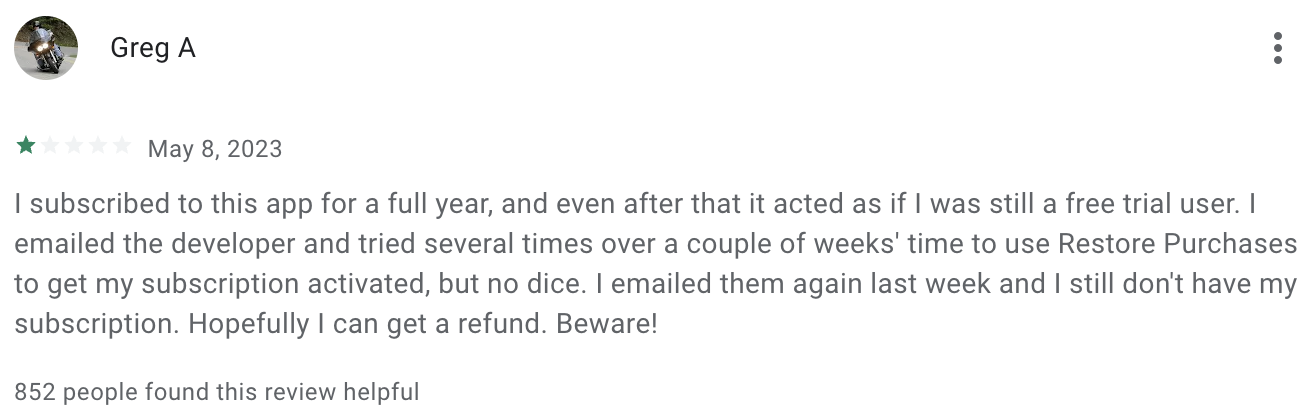}
    \caption{A user review for ``Ask AI - Chat with Chatbot" complaining about the subscription issue}
    \label{fig:chatgpt_review}
\end{figure}

\noindent\textbf{Zero-Day Apps}. Additionally, we discover $190$ apps uploaded to Google Play in February with no VirusTotal labels (i.e., this field is set to null) profiled by the AndroZoo dataset, acquiring label values by August.
Consequently, the security status of these apps was uncertain in February, leading us to categorize them as zero-day apps.
Within this group, $20$ apps have gained over one million downloads, and five have even more than $10$ million. 
Moreover, $64$ apps, constituting over one-third of the zero-day apps, have either fewer than $1,000$ downloads or have already been removed from the marketplace. 
Among the zero-day apps, we notice an uptick in chatbot apps leveraging GPT technology: $18$ are identified in our dataset, and five have crossed the one million downloads. 
Despite their popularity, user reviews reveal significant concerns about these chatbot apps, particularly regarding their premium services, which include issues like high costs, convoluted unsubscription procedures, and subpar customer service (see Fig.~\ref{fig:chatgpt_review}). 
Some are even labeled as scams or fleeceware by VirusTotal. 

\noindent\textbf{Late-Detection Malware}. Furthermore, we apply \sys's malware detection model on the new dataset generated by this temporal analysis.
\sys successfully identifies $28$ late-detection malware that were labeled as benign apps by VirusTotal in February, but later were reclassified as malware/PUAs in August.
For instance, \sys detects a barcode scanner malware, ``QR Scanner Rewards", promoted by several PUAs via ad libraries like Google AdMob and AppLovin. 
This malware exhibits behaviors identical to the malware that infected $10$ million users, as documented in a $2021$ security blog~\cite{malwarebytes_report}. 
Additionally, the PUA named ``Open Chat GBT - AI Chatbot App",  released in February, was marked as benign but subsequently labeled as fleeceware in August as user reports of its fraudulent premium service increased. 
Notably, this app is promoted by one of the ``lomol'' trojans,  which additionally promotes five malware and eight PUAs. 
\textit{These findings show that app promotion ads have become a fertile ground to distribute PUAs/malware.}

\subsection{RQ4: App Promotion Graph Construction}

\label{sec:eval-RQ1}

\begin{figure}[t]
    \centering
    \includegraphics[width=0.8\columnwidth]{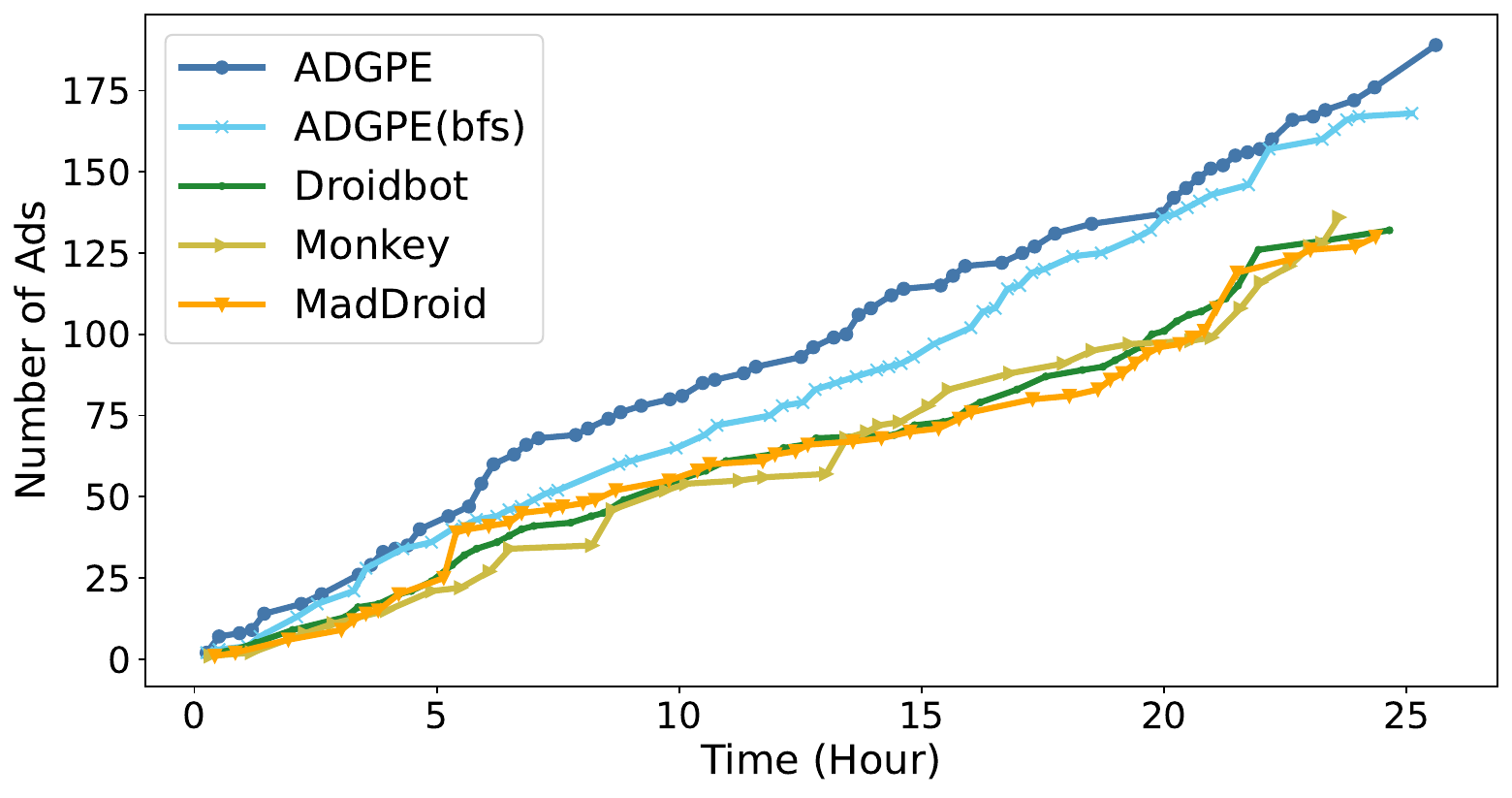}
    \caption{Unique ads collected by time}
    \label{fig:RQ1-efficiency}
\end{figure}

\begin{table}[t]
\centering
\caption{Comparison of ad coverage measured by collected ad units and ad types}
\label{tab:RQ1-ad_coverage}
\begin{tabular}{@{}lrrrr@{}}
\toprule
\multicolumn{1}{c}{\multirow{2}{*}{\textbf{Approaches}}} & \multicolumn{1}{c}{\multirow{2}{*}{\textbf{Ad Units}}} & \multicolumn{3}{c}{\textbf{Ad Types}}                                                  \\ \cmidrule(l){3-5} 
\multicolumn{1}{c}{}                                     & \multicolumn{1}{c}{}                                   & \multicolumn{1}{l}{Inherent} & \multicolumn{1}{l}{Pop-up} & \multicolumn{1}{l}{Custom-Made} \\ \midrule
Droidbot~\cite{li2017droidbot}                           & 76                                                     & 27                           & 38                         & 9                          \\
Monkey~\cite{monkey}                                     & 71                                                     & 26                           & 34                         & 11                         \\
\blue{DARPA~\cite{cai2023darpa}}                                & 8                                                      & 8                            & 0                          & 0                          \\
\blue{MadDroid~\cite{liu2020maddroid}}                          & 75                                                     & 32                           & 39                         & 6                          \\
\sys(bfs)                                                & 131                                                    & 52                           & 58                         & 15                         \\
\sys                                                     & \textbf{165}                                           & \textbf{76}                  & \textbf{71}                & \textbf{17}                \\ 
\bottomrule
\end{tabular}
\end{table}
\blue{To evaluate the efficiency of \sys in constructing the app promotion graph, we compare \sys with general-purpose UI exploration approaches such as Monkey~\cite{monkey}, DroidBot~\cite{li2017droidbot}, and \sys implemented with breadth-first strategy (coined as \sys(bfs)).
We also reproduce two SOTA ad detection approaches: MadDroid~\cite{liu2020maddroid} and DARPA~\cite{cai2023darpa}.
MadDroid uses breadth-first search to detect ads in WebView, ImageView, or ViewFlipper. 
DARPA leverages YOLO5~\cite{Jocher_YOLOv5_by_Ultralytics_2020} to detect encouraging patterns of ads.  Since DARPA lacks a UI exploration component, we manually labeled screenshots of app promotion ads from $30$ apps within our affordable efforts and applied DARPA to detect ads within these images.
Our comparisons use the same dataset from AndroZoo as the one used in Section~\ref{subsec:characteristics_ads}.
}

\noindent\textbf{Ad Units}.
We further evaluate the effectiveness of each approach in collecting ads based on the diversity of ad types and the total number of ad units. 
Ad units are placeholders for developers to show ads. 
The content within an ad unit can change over time, but the number of ad units in an app is determined by the source code and thus remains constant~\cite{admob_adformats}.
Additionally, we manually verify the number of ad units and the presence of each ad type in each app as ground truth. 

\blue{
The results in Table~\ref{tab:RQ1-ad_coverage} reveal that \sys
finds the most ad units ($165$), which achieves $117.1\%$ improvements over MadDroid, $132.4\%$ over Monkey, and $26.0\%$ over \sys(bfs).
Besides inherent ads and pop-up ads, \sys also is more effective in finding custom-made ads ($17$), which are at least $54.5\%$ better than MadDroid and Monkey.
}

\noindent\textbf{Efficiency}.
Fig.~\ref{fig:RQ1-efficiency} shows the number of unique app promotion ads collected over time by each approach.  
It can be seen that \sys significantly outperforms DroidBot and Monkey in collecting more ads with less time used, and performs slightly better than \sys(bfs). 
Specifically, \sys finds $54$ ads within $10$ hours, which is $20\%$ more than \sys(bfs) ($43$ ads), $37\%$ more than Droidbot ($34$ ads),  $24\%$ more than MadDroid, and $50\%$ more than Monkey ($27$ ads).

 \noindent\textbf{App Promotion Ads in the Wild}. %
We also conduct a case study on app promotion ads in the wild.
These ads direct users outside Google Play and thus we need to manually download the promoted apps.
Within our affordable efforts, we focus on a group of underground apps~\cite{chen2023deuedroid}, as their ads and corresponding download links are usually hardcoded and thus easy to detect. 
Specifically, we adopt the same strategy to build the app promotion graph and start from two seed apps. We obtain a graph of $37$ apps consisting of $21$ PUAs/malware ($5$ gambling, $11$ pornographic, $1$ trojan, $4$ adware).
Interestingly, we notice that a scam gambling app can be reached by both seed apps, potentially revealing an interconnected underground profit network.
\textit{This finding demonstrates the capability of \sys ad-oriented UI exploration technique in detecting app promotion ads in the wild and its usability in studying real-world problems such as the underground economy.}

\subsection{RQ5: App Promotion Reasoning}
\label{subsec:linkinfer}

\begin{table}[t]
\centering
\caption{Comparison of different link prediction models under two conditions}
\label{tab: model_evaluation}
\resizebox{\linewidth}{!}{
\begin{tabular}{@{}lrrrrrr@{}}
\toprule
\multirow{2}{*}{\textbf{Approaches}} & \multicolumn{3}{c}{\textbf{w/o pre-trained embedding}} & \multicolumn{3}{c}{\textbf{pre-trained embedding}} \\
& {Hit@$1$} & {  Hit@$10$} & {MRR} & {Hit@$1$} & {Hit@$10$} & {MRR} \\
\midrule
DistMult~\cite{DistMult} & $36.9$ & $61.1$ & $45.5$ & $40.9$ & $62.9$ & $48.6$  \\
Complex~\cite{complex} & $37.8$ & $60.0$ & $45.4$ & $33.8$ & $59.1$ & $42.4$ \\
Conve~\cite{conve} & $34.4$ & $56.1$ & $41.7$ & $36.1$ & $56.0$ & $42.6$ \\
\midrule
Path Inference & $\textbf{58.2}$ & $\textbf{66.6}$ & $\textbf{61.3}$ & $\textbf{59.5}$ & $\textbf{67.4}$ & $\textbf{62.4}$ \\
\bottomrule
\end{tabular}}
\end{table}

To evaluate the effectiveness of \sys in reasoning app promotion, we compare its performance with three SOTA link prediction models (DistMult~\cite{DistMult}, ComplEx~\cite{complex}, and ConvE~\cite{conve}) across two conditions (with and without the pre-trained graph embeddings). 
We employ two widely used metrics as follows:
\begin{itemize}[noitemsep, topsep=1pt, partopsep=1pt, listparindent=\parindent, leftmargin=*]
\item \textit{Hit@K} represents the proportion of times that the ground truth item appears in the top $K$ output of the model.
\item \textit{Mean Reciprocal Rank (MRR)} calculates the average of the reciprocal of the ranks of the ground truth items, with higher scores indicating better performance.
\end{itemize}

Table~\ref{tab: model_evaluation} shows the results of our evaluation.
First, we observe that the performance with pre-trained embedding is better than without pre-trained embedding, which shows that integrating app promotion relations into the entity representation captures more information about app promotion, thereby enabling better inference. 
Furthermore, Table~\ref{tab: model_evaluation} demonstrates that the path inference model outperforms the SOTA  embedding models, achieving a Hit@$1$ of $59.5\%$, a Hit@$10$ of $67.4\%$, and an MRR of $62.4\%$. 
Notably, the path inference model secures a lead of $45.48\%$ in Hit@$1$ and $28.40\%$ in MRR relative to the best-performing SOTA model (i.e., DistMult).

\noindent\textbf{Aiding UI Exploration}.
To evaluate the effectiveness of the path inference model in the wild, we use the model to predict the app promotion links in the test set, which are masked in the training process. 
In the total $907$ unique app promotion links, the model successfully predicts $318$ of them ($35.06\%$), indicating its potential to predict the promotion links that are not found by UI exploration.
To further evaluate the effectiveness of our prediction results in facilitating our UI exploration, we measure the effort required to identify the $318$ links.  
We do this by computing the average number of clicks made during the ad-oriented UI exploration using the recorded \textit{clickTrace} for each link. 
On average, the ad-oriented UI exploration requires $18.3$ clicks to identify a predicted link.
Surprisingly, \sys's path inference model can even predict a link about an appwall ad by Google AdMob that requires $54$ clicks to obtain. 
Given that the inference time of the path inference model is negligible compared to the resources and the time spent on UI exploration, these results demonstrate the model's effectiveness in augmenting \sys's ad-oriented UI exploration.
\textit{This demonstrates the promising direction of integrating AI to aid program analysis, consistent with previous work}~\cite{liu2022promal,deepintent,adamo2018reinforcement,hu2011automating}.

  \begin{figure}[t]
    \centering
    \begin{subfigure}[]{0.80\linewidth}
      \centering
      \includegraphics[width=\linewidth]{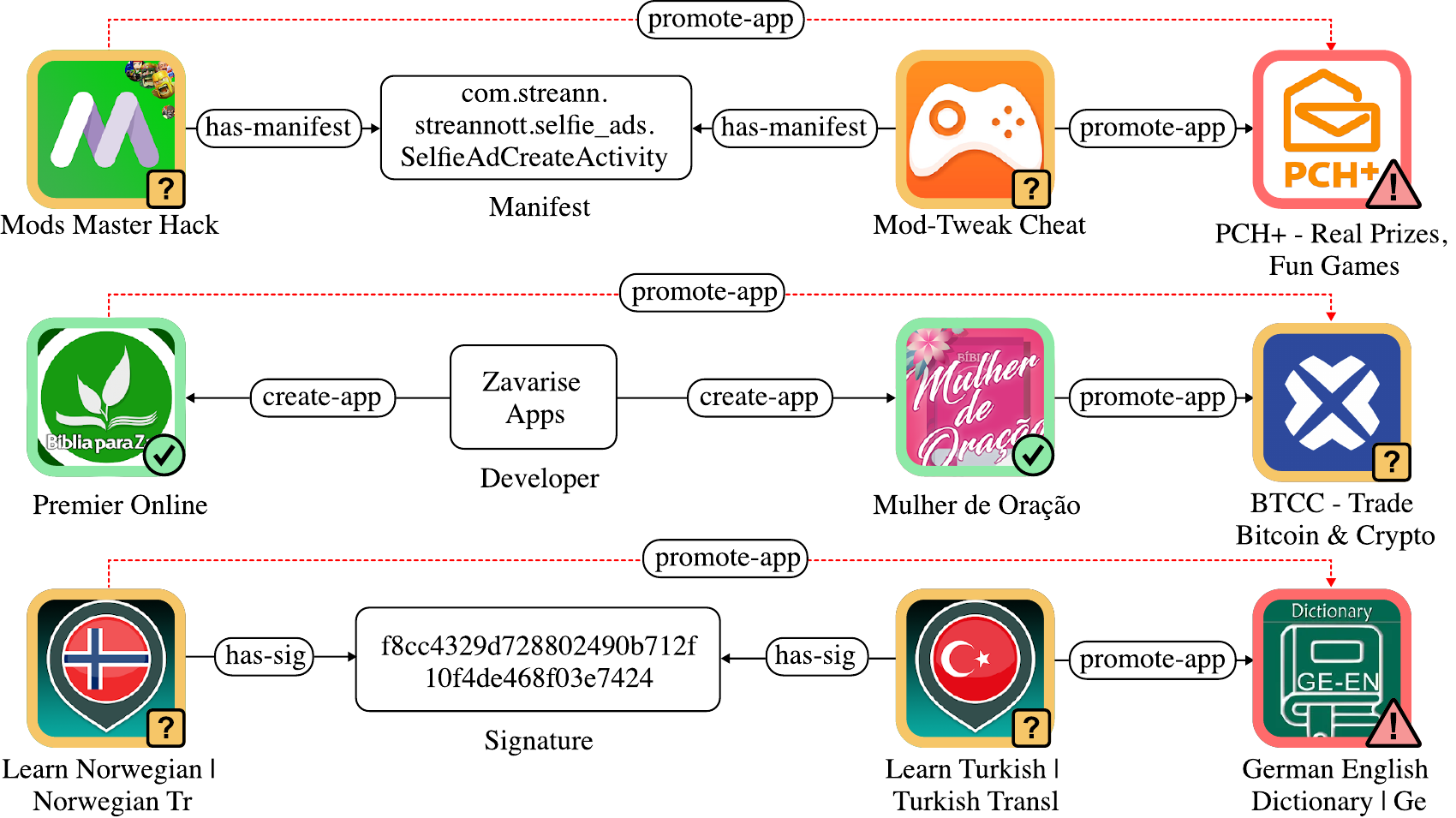}
      \caption{Custom-made ad-based propagation}
      \label{fig:channel_shared_codes}
    \end{subfigure}
    \vspace{0.01em}
    \begin{subfigure}[]{0.80\linewidth}
      \centering
      \includegraphics[width=\linewidth]{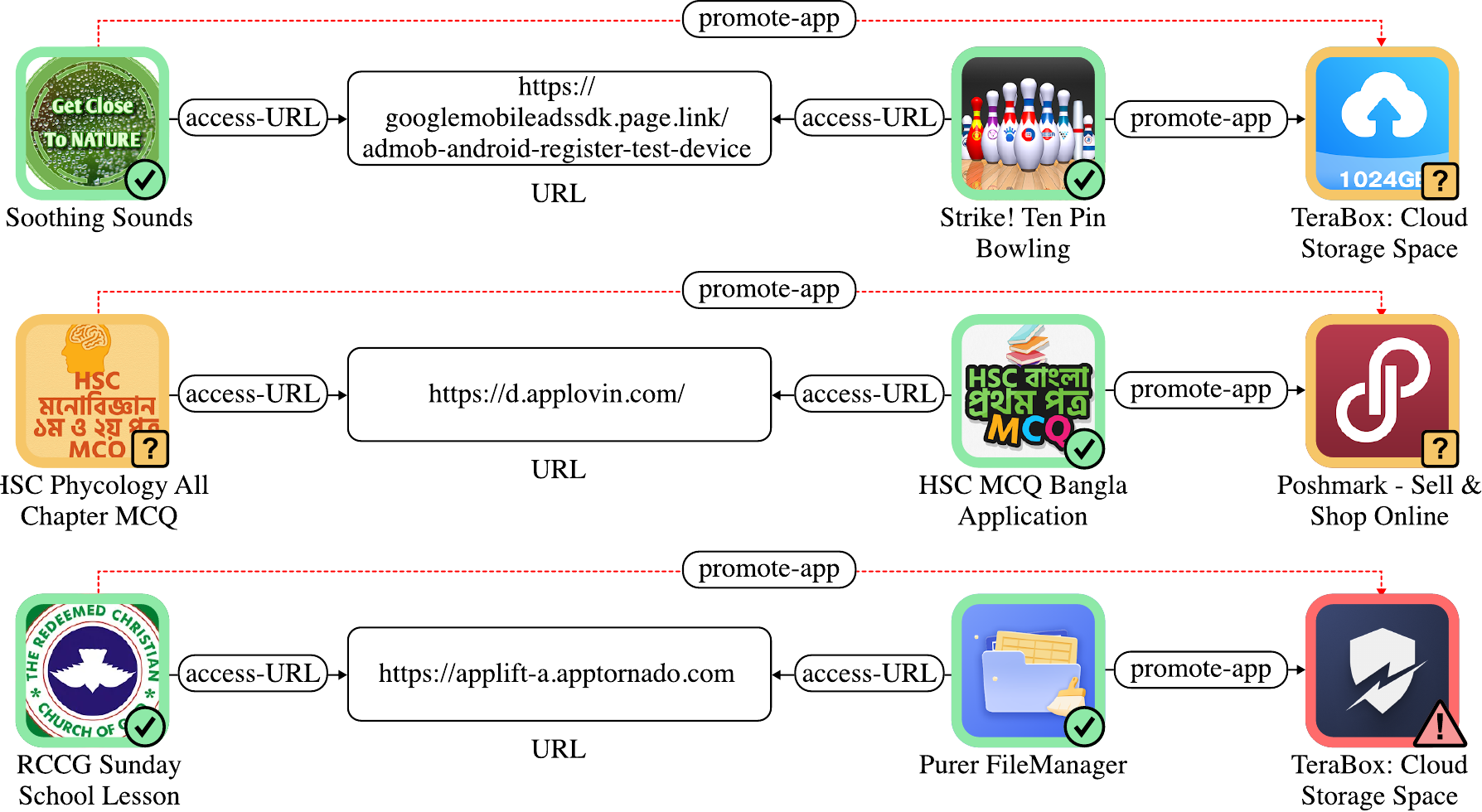}
      \caption{Ad library-based propagation}
      \label{fig:channel_ad_lib}
    \end{subfigure}
    \caption{Examples of malware/PUA propagation mechanisms}
    \label{fig:propagation_channels}
  \end{figure}

\noindent\textbf{Uncovering Malware Promotion Mechanisms}.
We examine the inference paths and uncover two typical malware promotion mechanisms: \textit{custom-made ad-based promotion mechanism}, which refers to app promotion ads hardcoded in the app's program, and \textit{ad library-based promotion mechanism}, which relates to the app's interactions with ad servers. 
Fig.~\ref{fig:propagation_channels} illustrates these promotion mechanisms.

The custom-made ad-based promotion mechanism stems from shared developers, signatures, or manifest components among multiple apps. Such common factors lead the apps to exhibit similar code structures that tend to promote similar apps through app promotion ads. For instance, our model infers that a language learning app ``Learn Norwegian-Norwegian Tr'' promotes malware ``German English Dictionary-Ge'' because both ``Learn Norwegian-Norwegian Tr'' and ``Learn Turkish-Turkish Transl'', another language learning app, share the same signature and are developed by the same developer ``lomol translator''. As a result, both language learning apps share the same code structure related to app promotion and are inclined to promote the same apps.  

The ad library-based promotion mechanism stems from the common URLs associated with ad libraries (e.g., AdMob) or ad promotion companies (e.g., Applovin) that are accessed by the apps.
For example, a Sunday school learning app ``RCCG Sunday School Lesson'' and a file management app ``Purer FileManager'' access the same URL related to an ad promotion company AppTornado. Meanwhile, the ``Purer FileManager'' promotes a malicious cloud storage app ``TeraBox: Cloud Storage Space.''
Based on such observation, the model concludes that ``RCCG Sunday School Lesson'' is highly likely to promote ``TeraBox: Cloud Storage Space''. This network promotion mechanism reveals the interconnected nature of app promotion graphs, as different apps would access the same ad server through common URLs.
These findings indicate that the path inference model is effective in explaining the reason for malware promotion, thereby enhancing the explainability for malware detection.

\blue{Through analyzing the promotion paths of $950$ verified malware/PUA in PIG, we found that the majority of advertiser apps were in the education ($27.7\%$) and books ($14.6\%$) categories. The majority of promoted malware/PUA apps were in the education ($12.6\%$), entertainment ($11.4\%$), and books ($11.4\%$) categories. Among all these promotion paths, 697 paths ($73.4\%$) used custom-made ads for promotion, while 253 paths ($26.6\%$) used ad libraries for promotion.}

\section{Discussion}

\noindent\textbf{Implications}.
Our findings reveal a concerning result that many benign apps are directly or indirectly involved in the promotion of malware, indicating inadequate vetting performed by ad libraries and app markets.
To address this, we advocate more rigorous vetting processes, stricter developer policies, and the adoption of open-source ad libraries~\cite{EthicalAdServer}. 
Meanwhile, users should exercise caution and if necessary, employ security tools to vet the recommended apps before installation.

\noindent\textbf{Limitations}.
The limitations of our app promotion graph construction are two-fold. 
First, our technique struggles with ads that appear after complex human interactions, such as ads that appear upon game failure~\cite{Xu2020}. We partially address this by expanding our seed app dataset. Second, false positives may arise, for instance, when a user is directed to Google Play for non-ad-related reasons like clicking a Facebook sharing button when Facebook is not installed.
The main limitation of our path inference model is its lack of consideration for ad targeting, which some apps use to deliver personalized ads. While integrating personalized data like location and device information into our model could enhance its ability to detect and analyze such ads, emulating user profiles is a complex task and would require separate research efforts~\cite{book2015empirical,nath2015madscope}.

\noindent\textbf{Future Works}.
Our future works are three-fold.
First, we plan to develop a unified mobile ads testing tool that uses static analysis~\cite{xiao2019iconintent,rountev2014static} to guide the UI exploration technique.
This technique will further leverage large language models~\cite{chatgpt,chatgpt2,large_language_models} to interact with the ads that require complex user interaction. 
Second, we plan to establish a long-term monitoring system, which allows us to track the changes and emerging trends in the app promotion ecosystem.
This will also facilitate the app community to develop more effective countermeasures against potential threats.
\blue{Additionally, we plan to explore the Chinese app promotion ecosystem which is distinct due to the absence of an official app market,  reliance on its own ad libraries, and the common practice of directing users to developer-specific websites for app downloads.}

\section{Related Work}



\noindent\textbf{Mobile Advertisements Identification.}
Previous research on mobile ads identification can be categorized into two types: identifying ad libraries, and identifying ad contents.
For ad libraries identification, many studies compared the package names with a whitelist of collected ad libraries~\cite{grace2012unsafe, book2013longitudinal, chen2014achieving}. 
Recent works focused on using clustering-based methods to detect such libraries and achieve very high accuracy~\cite{ma2016libradar, li2017libd}. 
However, ad content is dynamically generated and cannot be obtained via static analysis.
Dynamic testing for UI exploration is a major approach for ad content identification~\cite{hu2011automating,rastogi2013appsplayground,nath2013smartads,liu2014decaf,hao2014puma}. 
For example, MadDroid combined breadth-first UI exploration with rule-based HTTP hooking to extract ad content~\cite{liu2020maddroid}. Researchers also leveraged computer vision techniques to detect button edges~\cite{rastogi2016these} or ad-related visual patterns~\cite{cai2023darpa} to recognize ad content.
However, these techniques are biased to specific ad types, resulting low effectiveness in detecting app promotion ads.
\sys gains insights from both our preliminary study and the design of existing techniques~\cite{liu2020maddroid,rastogi2016these,cai2023darpa}.

\noindent\textbf{Malicious Advertisements Detection.}
\blue{The most pertinent previous research to our work is malicious mobile advertising.
Existing studies analyzed deceptive ad contents~\cite{liu2020maddroid} and malicious destinations (e.g., files, apps, and external websites) triggered by mobile ads~\cite{rastogi2016these}. 
However, these studies mainly examined the malicious behaviors of ad libraries, and none of them have specifically studied malware promotions through mobile ads.
Another line of related work is graph-based Android malware detection. 
GNNs have gained prominence in this area due to their strong capabilities in representing structural data.  
APIGraph employed official Android API documentation to construct a relation graph and utilizes TransE~\cite{bordes2013translating} to learn the graph embeddings, which are subsequently used for API clustering and malware detection~\cite{zhang2020enhancing}. 
Both HinDroid~\cite{hou2017hindroid} and Hawk~\cite{hei2021hawk} constructed a heterogeneous information network to model the relationship of malware based on API calls and app attributes.
\sys diverges from these approaches by incorporating the app promotion relations. Moreover, the embedding trained by \sys serves more downstream tasks, such as promotion path inference.
}


\noindent\textbf{Graph Path Inference. }
Reinforcement learning~\cite{kaelbling1996reinforcement} (RL) is commonly used for path inference in graphs, treating the task as a Markov decision process.
DeepPath~\cite{williams1992simple} was the first RL-based model to find representative paths between entity pairs and employed a path ranking algorithm for training.
Furthermore, MINERVA~\cite{dasgo2018ICLR} learned how to guide the graph depending on the input entity-relation pair to find predictive paths where the second entity is unknown and must be acquired by inferring.
Multi-Hop~\cite{lin2018multi} proposed two improvements over MINERVA: reward shaping which uses soft rewards to capture triple semantics, and action dropout which enables more effective path exploration.
We build our path inference model upon Multi-Hop to infer app promotion paths and explain malware promotion mechanisms. 

\section{Conclusion}
We introduce a novel approach, \sys, that synergistically integrates app UI exploration with graph learning to automatically collect app promotion ads, detect malware promoted by these ads, and explain the malware promotion mechanisms employed by the detected malware. 
\blue{Our analysis of $18,627$ app promotion ads reveals a heightened risk for downloading apps via app promotion ads, which is hundreds of times higher than the likelihood of downloading malware from Google Play. Popular ad libraries are exploited by malicious developers to distribute a variety of malware. Our evaluations on real apps show that our malware detection model outperforms existing models in detecting ad-promoted malware, and our path inference model further reveals two primary malware promotion mechanisms. 
These findings demonstrate the effectiveness of combining dynamic program analysis with graph learning in studying malware promotion.}

\section*{Acknowledgments}
Xusheng Xiao's work is partially supported by the National Science Foundation under the grant CCF-2318483. Toby Jia-jun Li's work was supported in part by an AnalytiXIN Faculty Fellowship, an NVIDIA Academic Hardware Grant, a Google Cloud Research Credit Award, a Google Research Scholar Award, a Google PSS Privacy Research Award, NSF Grant 2341187, and NSF Grant 2326378. Any opinions, findings, and conclusions or recommendations expressed in this material are those of the authors and do not necessarily reflect the views of the sponsors.



\bibliographystyle{IEEEtran}
%

\bibliography{Mybib}


\end{document}